\documentclass[letterpaper]{article} 
\usepackage{aaai24}  
\usepackage{times}  
\usepackage{helvet}  
\usepackage{courier}  
\usepackage[hyphens]{url}  
\usepackage{graphicx} 
\usepackage{natbib}  
\usepackage{caption} 
\usepackage{algorithm}
\usepackage{algorithmic}

\usepackage{newfloat}
\usepackage{listings}
\DeclareCaptionStyle{ruled}{labelfont=normalfont,labelsep=colon,strut=off} 
\floatstyle{ruled}
\newfloat{listing}{tb}{lst}{}
\floatname{listing}{Listing}
\usepackage{amsmath}
\usepackage{mathtools}
\usepackage{bm}
\usepackage{multirow}
\usepackage{booktabs}
\usepackage{subcaption}

\newcommand\tool{\textsc{DataElixir}}
\def\eg{\emph{e.g.}}
\def\etc{\emph{etc.}}

\title{\tool: Purifying Poisoned Dataset to Mitigate Backdoor Attacks \\ via Diffusion Models}
\author{
    Jiachen Zhou\textsuperscript{\rm 1,\rm 2}, Peizhuo Lv\textsuperscript{\rm 1,\rm 2}, Yibing Lan\textsuperscript{\rm 1,\rm 2}, Guozhu Meng\textsuperscript{\rm 1,\rm 2}\thanks{Corresponding Authors.}, Kai Chen\textsuperscript{\rm 1,\rm 2}\footnotemark[1], Hualong Ma\textsuperscript{\rm 1,\rm 2}\\
}
\affiliations{
    \textsuperscript{\rm 1}Institute of Information Engineering, Chinese Academy of Sciences, China\\
    \textsuperscript{\rm 2}School of Cyber Security, University of Chinese Academy of Sciences, China
    
    \{zhoujiachen, lvpeizhuo, lanyibing, mengguozhu, chenkai, mahualong\}@iie.ac.cn
}

\begin{document}

\maketitle

\begin{abstract}
Dataset sanitization is a widely adopted proactive defense against poisoning-based backdoor attacks, aimed at filtering out and removing poisoned samples from training datasets. However, existing methods have shown limited efficacy in countering the ever-evolving trigger functions, and often leading to considerable degradation of benign accuracy. In this paper, we propose \tool, a novel sanitization approach tailored to purify poisoned datasets. We leverage diffusion models to eliminate trigger features and restore benign features, thereby turning the poisoned samples into benign ones. Specifically, with multiple iterations of the forward and reverse process, we extract intermediary images and their predicted labels for each sample in the original dataset. Then, we identify anomalous samples in terms of the presence of label transition of the intermediary images, detect the target label by quantifying distribution discrepancy, select their purified images considering pixel and feature distance, and determine their ground-truth labels by training a benign model. Experiments conducted on 9 popular attacks demonstrates that \tool~effectively mitigates various complex attacks while exerting minimal impact on benign accuracy, surpassing the performance of baseline defense methods.
\end{abstract}

\section{Introduction}
\label{sec:introduction}

Deep neural networks (DNNs) have exhibited remarkable performance in various fields, including autonomous driving~\cite{grigorescu2020survey}, facial recognition~\cite{wang2021deep} and medical image analysis~\cite{shen2017deep},~\etc~Advanced DNN models necessitate large and diverse high-quality data for training. However, collecting high-quality data is an exceedingly resource-intensive task. For instance, ImageNet~\cite{deng2009imagenet} took more than two years to construct and consumed substantial resources. Therefore, it has become customary for individuals to leverage data from external sources, such as data markets and crowd-sourcing platforms~\cite{zeng2023meta}. 

However, except for a few authoritative and community-maintained sources that are well maintained and thereby have high-quality data~\cite{huggingface2016datasets}, other sources may be poisoned with backdoor samples by miscreants~\cite{li2022backdoor}. In particular, backdoor attacks craft poisoned samples by using trigger functions to modify benign images. Victim models trained on these poisoned datasets, although can make accurate predictions for benign inputs, will be controlled to produce wrong outputs given poisoned inputs. 

\begin{figure}[!t]
   \centering
    \includegraphics[width=0.95\columnwidth]{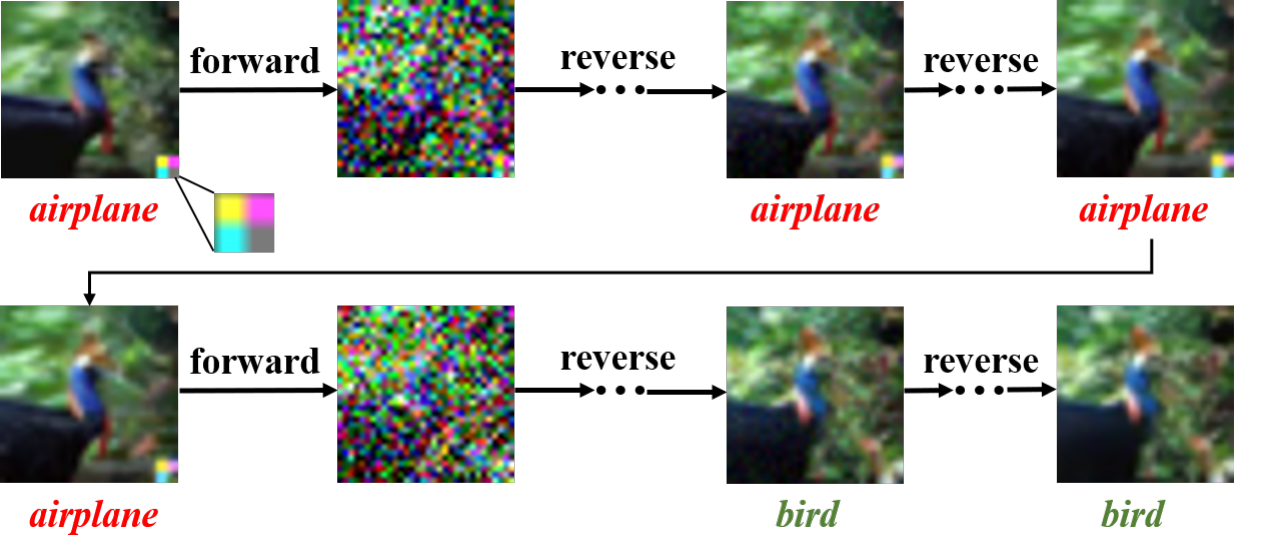}
    \caption{The forward and reverse process in diffusion models. The poisoned image is misclassified as ``airplane'' due to the trigger in the lower right. The forward process eliminates this trigger by introducing noise, while the reverse process restores the benign features, correcting the model to its ground-truth label ``bird''.}
    \label{fig1}
\end{figure}

A straightforward and effective way to defend such backdoor attacks is to identify in the dataset the poisoned samples and further remove or replace them. Prior studies propose dataset sanitization methods to identify poisoned samples by detecting their anomalies in model activations~\cite{chen2019detecting}, spectral signatures~\cite{tran2018spectral}, and loss values~\cite{li2021anti,huang2022backdoor}. Subsequently, they remove identified anomalous samples from the dataset to train benign models free from backdoors. However, existing methods have exhibited limited effectiveness, only working on specific types of trigger functions and often resulting in considerable degradation of benign accuracy~\cite{li2022backdoor,wu2022backdoorbench}. Similar challenges are also encountered in other types of defense approaches, including model reconstruction~\cite{wang2019neural,wu2021adversarial,li2021neural,zeng2022adversarial} and input purification~\cite{doan2020februus,shi2023blackbox,may2023salient}.

In this paper, we propose \tool, a novel sanitization approach tailored to purify poisoned datasets. We utilize the forward and reverse process of diffusion models to turn poisoned samples into benign ones, thus training benign models devoid of backdoors. As shown in Figure~\ref{fig1}, the forward process introduces noise to the image, thus eliminating the trigger features and preventing the victim model from making predictions based on them. While the reverse process restores the benign features of the image, making the victim model to reclassify it as its ground-truth label.

Specifically, for each sample in the training dataset, we perform iterative rounds through the forward and reverse process and extract intermediary images from each round to construct the candidate set. Based on the presence of label transition in their candidate sets, we categorize samples into three types: the Benign Set, Poisoned Set and Suspicious Set, of which the latter two are regarded as anomalous. Then we detect the target label by quantifying its distribution discrepancy in the candidate sets of anomalous samples. Ultimately, for these anomalous samples, we select their purified images from their respective candidate sets considering pixel and feature distance, and determine their ground-truth labels through the training of a benign model. 

We evaluate the efficacy of \tool~against 9 popular backdoor attacks. Compared to 4 baseline defense methods aimed at dataset sanitation, \tool~demonstrates superior performance by effectively mitigating diverse trigger functions, while maintaining the benign accuracy of the model. In particular, in regard to the detection rate of poisoned samples, our approach achieves considerably increases spanning from 37.02\% to 56.77\% on CIFAR10 and from 1.85\% to 62.10\% on Tiny ImageNet, as well as the decrease in false positive rates from 2.11\% to 47.46\% on CIFAR10 and from 0.38\% to 35.60\% on Tiny ImageNet. Moreover, the benign models obtained by \tool~also exhibit remarkable enhancements, with an average increase in accuracy from 2.81\% to 17.27\% on CIFAR10 and from 0.20\% to 36.90\% on Tiny ImageNet, and the decrease in backdoor attack success rate from 10.73\% to 84.81\% on CIFAR10 and from 8.45\% to 97.56\% on Tiny ImageNet. 

Our contributions are summarized as follows:
\begin{itemize}
    \item We propose \tool\footnote{\url{https://github.com/Manu21JC/DataElixir}}, which is able to accurately identify poisoned samples as well as their target labels without prior knowledge of trigger functions and patterns. The poisoned samples can be further purified and replaced with the benign through diffusion models.
    \item Experiments against 9 popular attacks highlight the effectiveness and generalizability of \tool~in mitigating backdoors, while preserving the benign accuracy, and outperforming mainstream defense methods.
\end{itemize}

\section{Related Work}
\label{sec:related work} 

\subsection{Diffusion Model}
\label{subsec:diffusion model}

Diffusion models~\cite{ho2020denoising,nichol2021improved,rombach2022high} have exhibited excellent performance in generating diverse and high-quality images. The original DDPM consists of two Markov Chains, wherein the forward process adds noise to the image, and the reverse process restores the image from Gaussian noise using neural networks. Recent studies~\cite{yoon2021adversarial,nie2022diffusion,xiao2023densepure,carlini2023certified} have attempted to employ diffusion models for adversarial purification. Different from these methods primarily target the purification of adversarial perturbation in inputs during the inference stage, \tool~harnesses the potential of diffusion models to counter backdoor triggers, and aims at purifying datasets poisoned by backdoor attacks to serve for training benign models.  

\subsection{Backdoor Attack}
\label{subsec:backdoor attack}

Backdoor attacks aim to induce the victim model to accurately predict labels for benign inputs, while intentionally misclassifying poisoned inputs to the target label. Among various techniques of backdoor attacks, poisoning the training datasets is the most common method. Attackers craft poisoned samples by employing trigger functions to modify benign images and altering their labels to the target one. Since the pioneering backdoor attack BadNets~\cite{gu2019badnets} that simply stamped patterns (\eg, 3 × 3 white square) on benign images, various trigger functions have been proposed to enhance the attack effectiveness and imperceptibility, including Invisible triggers~\cite{chen2017targeted,li2020invisible}, Clean Label triggers~\cite{shafahi2018poison}, Sample Specific triggers~\cite{nguyen2020input,nguyen2021wanet,doan2021lira,li2021invisible}, and Frequency triggers~\cite{zeng2021rethinking,wang2022invisible},~\etc

Besides poisoning-based backdoor attacks, there exist other methods to embed backdoors without poisoning training data, including manipulating model weights~\cite{garg2020can,rakin2020tbt} or altering model structures~\cite{tang2020embarrassingly,li2021deeppayload}. The primary focus of our study is on purifying the poisoned training datasets to defend backdoor attacks as 1) data poisoning is the most generalized way to implant backdoors due to its agnosticism to model structures. 2) it takes only one-time effort but can benefit an unlimited number of model training tasks.

\subsection{Backdoor Defense}
\label{subsec:backdoor defense}

To defend against poisoning-based backdoor attacks, several approaches have been proposed to sanitize the poisoned datasets. These methods focus on filtering out poisoned samples based on their anomalies in model activations~\cite{chen2019detecting}, spectral signatures~\cite{tran2018spectral}, and loss values~\cite{li2021anti,huang2022backdoor}. Then identified samples are removed to avoid potential injection of backdoors and enable the training of benign models. 

Except for dataset sanitation, model reconstruction aims to mitigate the impact of backdoors in the infected model by pruning poisoned neurons~\cite{wu2021adversarial}, fine-tuning model weights~\cite{li2021neural,zeng2022adversarial}, and reversing backdoor triggers~\cite{wang2019neural}. While input purification seeks to purify poisoned inputs using generative adversarial networks~\cite{doan2020februus,cheng2023beagle} and diffusion models~\cite{shi2023blackbox,may2023salient} during the inference stage, thus preventing them to activate backdoors in the infected model.

Unlike previous dataset sanitization methods, our approach aims to purify identified poisoned samples into benign instances instead of outright removal, including the determination of their purified images and ground-truth labels. While some input purification methods also utilize diffusion models to eliminate the impact of triggers, \tool~is different from them as 1) the purification of inputs occurs during the inference stage, aimed at preventing the activation of backdoors in the infected model rather than clearing them from existence. 2) these methods focus solely on image purification, while \tool~encompasses a broader scope, including identifying the poisoned samples and determining their ground-truth labels.

\section{Methodology}
\label{sec:methodology}

\begin{figure*}[!t]
\centering
\includegraphics[width=2.1\columnwidth]{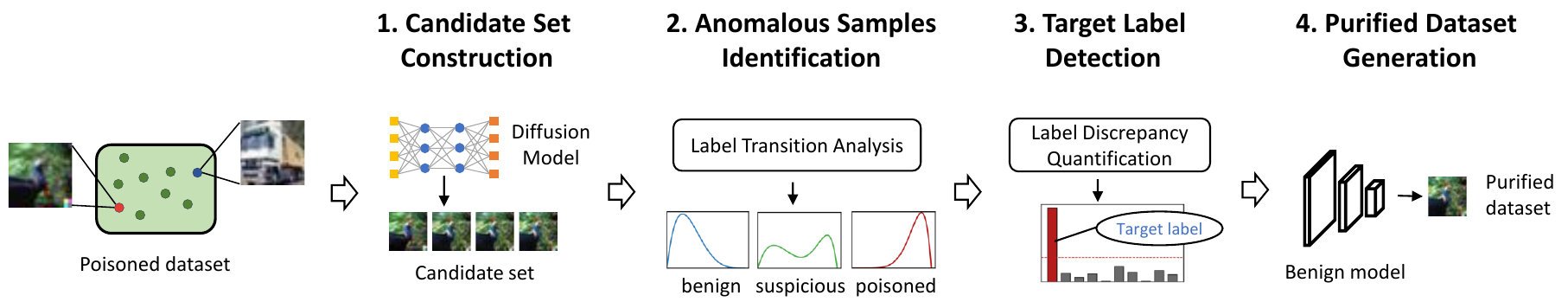}
\caption{Overview of \tool.}
\label{fig2}
\end{figure*}

In this section, we present the threat model and methodology of our proposed approach. As shown in Figure~\ref{fig2}, \tool~consists of four stages: Candidate Set Construction, Anomalous Samples Identification, Target Label Detection and Purified Dataset Generation. Comprehensive technical details are introduced in the following subsections.

\subsection{Threat Model}
\label{subsec:threat model}

Given the untrustworthy data from external sources, the defender needs to determine whether it is poisoned, and if poisoned, identify and purify the poisoned samples for training benign models free from backdoors. This defense mechanism operates in a black-box manner, without prior knowledge of the attack settings (\eg, the target label, poison rate and trigger functions), as well as any known benign or poisoned samples from the given data. We assume that the defender has access to pre-trained diffusion models representing identical or similar distributions (consistent with prior studies~\cite{nie2022diffusion,xiao2023densepure,may2023salient,shi2023blackbox,dolatabadi2023devils,jiang2023unlearnable}), which can be sourced as off-the-shelf models or trained using data provided from authoritative and trustworthy communities and institutions. Beyond this assumption, our experiments also validate the effectiveness of \tool~using diffusion models trained on disparate data (Section~\ref{subsec:ablation study}) and even poisoned data (Section~\ref{subsec:adaptive attack}).

\subsection{Candidate Set Construction}
\label{subsec:candidate set construction}

The raw dataset $D=\{(x_i, y_i)\}_{i=1}^N$ may contain meticulously crafted poisoned samples that inject backdoors into the victim model $M$ trained on it. As shown in Figure~\ref{fig1}, we utilize the forward and reverse process of diffusion models for the purification of poisoned samples. Specifically, the forward process $q$ adds Gaussian noise to the original image $x^0$ for $T$ steps (from $x^0$ to $x^T$) according to a variance schedule $\{\beta^t\in(0,1)\}^T_{t=1}$, yielding the noised image $x^T$:  
\begin{equation}
\label{eq1}
\begin{gathered}	   
q(x^{1:T}|x^0)\coloneqq\prod^{T}_{t=1}q(x^t|x^{t-1}) \\ q(x^t|x^{t-1})\coloneqq\mathcal{N}(x^t;\sqrt{1-\beta^t}x^{t-1},\beta^t\bm{I})
\end{gathered}
\end{equation}

While training on the poisoned dataset, the model $M$ is extremely overfitting to the trigger features. This prompts $M$ to predict the poisoned image as the target label, instead of its ground-truth label based on its benign features. During the forward process, the introduction of noise eliminates these trigger features thus would change the prediction of $M$. However, the benign features are also degraded by the noise. We use the reverse process $p_{\theta}$ to restore benign features (from $x^T$ back to $x^0$), yielding the purified image $\widetilde{x}$:
\begin{equation}
\label{eq2}
\begin{gathered}	   
    p_{\theta}(x^{0:T})\coloneqq p(x^T)\prod^{T}_{t=1}p_{\theta}(x^{t-1}|x^t) \\ p_{\theta}(x^{t-1}|x^t)\coloneqq\mathcal{N}(x^{t-1};\mu_{\theta}(x^t,t),\Sigma_\theta(x^t,t))
\end{gathered}
\end{equation}

The mean $\mu_{\theta}(x^t,t)$ is learned by a neural network parameterized by $\theta$, and the variance $\Sigma_\theta(x^t,t)$ can either be time-step dependent constants~\cite{ho2020denoising} or learned by a neural network~\cite{nichol2021improved}. The reverse process restores the image with guidance from the remaining benign features. Since the diffusion model represents the benign distribution, the benign features degraded during the forward process can be well-recovered whereas the trigger features cannot. Consequently, the purified image $\widetilde{x}$ only contains the benign features of $x$, thus making $M$ correctly classify it as its ground-truth label.  

During the purification of each image, to effectively eliminate all trigger features, we iterate through the forward and reverse process for $n$ rounds, where the reverse process of each round consists of $T$ steps, thereby producing $n \times T$ intermediary images. If an image is poisoned, the labels of its intermediary images transit from the target label back to the ground-truth label as the trigger features are eliminated. In contrast, the purification of the benign image does not involve such label transition. Thus, the presence of transition in the labels of the intermediary images provides a strong indication that the sample is likely poisoned. Specifically, for the $i$th sample $(x_i,y_i)$ in $D$, we perform the forward and reverse process for $n$ rounds, and extract intermediary images from the last $m$ steps in the reverse process of each round to construct its candidate set $C_{(x_i,y_i)}=\{(x_j, y_j)\}_{j=1}^{n \times m}$ . We can then determine whether the sample $(x_i,y_i)$ is anomalous based on the presence of label transition in $C_{(x_i,y_i)}$.

\subsection{Anomalous Samples Identification}
\label{subsec:anomalous samples identification}

\noindent\textbf{Label Transition Analysis.} When assessing the presence of label transition in the candidate set, we consider the second-highest count of labels as the transition coefficient $\eta$. If $\eta$ surpasses the predefined threshold $\tau$, we consider it indicative of label transition. The motivation behind $\eta$ is as follows. Ideally, the label distribution of the candidate set for benign sample should exhibit a unimodal distribution, wherein all labels align with the ground-truth label. While the label distribution transits to a bimodal distribution with $\eta \ge \tau$, it indicates that the sample is poisoned. The appearance of the second peak in the distribution is most likely due to the transition from the target label to the ground-truth label, a direct result of the effective elimination of trigger features.

We categorize samples in the raw dataset $D$ into three types based on the presence of label transition in the candidate set: the Benign Set $B$, Poisoned Set $P$ and Suspicious Set $S$, where samples in $P$ and $S$ are considered anomalous:

\begin{equation}
\small
\label{eq3}
  \begin{gathered}
    (x_i, y_i) \in
    \begin{cases}
      B, &\text{if}\ \eta < \tau, \forall (x_j,y_j) \in C_{(x_i, y_i)}, y_j=y_i \\
      P, &\text{if}\ \eta < \tau, \forall (x_j,y_j) \in C_{(x_i, y_i)}, y_j=y_g \\
      S, &\text{if}\ \eta \ge \tau
    \end{cases} \\
    B \cup P \cup S = D
  \end{gathered}
\end{equation}

In particular, for the sample $(x_i,y_i)$ where $\eta < \tau$, if the labels of $C_{(x_i,y_i)}$ remain consistent with $y_i$, this provides conclusive evidence that the sample is benign, with its benign features remaining intact throughout the entire purification process. Such samples constitute the Benign Set $B$, with $x_i$ as the purified image and $y_i$ as the ground-truth label, as neither have been poisoned or modified. 

Conversely, the Poisoned Set $P$ also comprises samples where $\eta < \tau$, but the labels of $C_{(x_i,y_i)}$ deviate from $y_i$, indicating that these samples are poisoned, with trigger features effectively eliminated during the initial round of forward and reverse process and then classified as their ground-truth label $y_g$. For such samples, we need to select their purified images from their candidate set since $x_i$ has been poisoned, and $y_g$ can be directly used as the ground-truth label. 

The Suspicious Set $S$ comprises samples where $\eta \ge \tau$, which can arise in two cases. Firstly, trigger features on the poisoned image may not be effectively eliminated until later iterations, causing the labels to exhibit the transition from the target label to the ground-truth label. Secondly, benign features on the benign image may be excessively corrupted, leading to false positives. For samples in $S$, we need to further select their purified images from their candidate set and determine their ground-truth label in Section~\ref{subsec:purified dataset generation}.

\subsection{Target Label Detection}
\label{subsec:target label detection}

Upon identifying the anomalous samples in Poisoned Set $P$ and Suspicious Set $S$, we proceed to determine whether the dataset is poisoned by detecting the target label. If any outlier label is detected, it indicates that the dataset has been poisoned, with this label corresponding to the target label. This detection process leverages a crucial insight: there exists distribution discrepancy between samples with the target label and those with benign labels in $P \cup S$. This discrepancy arises because the samples with the target label in $P \cup S$ mainly consists of poisoned samples, while samples with benign labels are primarily false positives of benign samples. We utilize the candidate sets of samples with each label to characterize such distribution discrepancy, followed by employing outlier detection methods such as \textit{Median Absolute Deviation}~\cite{leys2013detecting} to detect the target label.

\noindent\textbf{Label Discrepancy Analysis.} Specifically, for each label $\mathrm{y}$, we construct $\mathcal{C}_\mathrm{y}$ as the union of the candidate sets $C_{(x_i,y_i)}$ for samples $(x_i,y_i)$ in $P \cup S$ with label $\mathrm{y}$ as follows: 
\begin{equation}
\label{eq4}
\begin{gathered}	   
    \mathcal{C}_\mathrm{y} = \bigcup\limits_{(x_i,y_i) \in P \cup S}\{C_{(x_i,y_i)} \mid y_i={\small \mathrm{y}}\}
\end{gathered}
\end{equation}

We provide two metrics to measure label discrepancy for the candidate set. Firstly, we count the number of samples in $\mathcal{C}_\mathrm{y}$, as the count of false positives of benign samples is lower than the count of poisoned samples. Secondly, we compute the entropy of the label distribution in $\mathcal{C}_\mathrm{y}$. Since the poisoned samples originate from many other benign labels, during the purification process, they are reclassified to their ground-truth labels due to the effective removal of trigger features. Therefore, the label distribution in $\mathcal{C}_\mathrm{y}$ for the target label is more chaotic than that of benign labels, where the samples are mere false positives of benign samples. For $\mathcal{C}_\mathrm{y}$ of each label, we calculate the \textit{Kullback-Leibler divergence}~\cite{kullback1951information} between the actual distribution and the ideal distribution where all labels are assumed to be identical. A higher value of \textit{KL divergence} indicates a higher likelihood of the label being the target label.

\subsection{Purified Dataset Generation}
\label{subsec:purified dataset generation}

We can generate the purified dataset $\widetilde{D}=\{(\widetilde{x_i}, \widetilde{y_i})\}_{i=1}^N$. For each sample $(x_i,y_i)$ with benign labels in $B \cup P\cup S$ and with the target label in $B$, it can be directly used as the purified sample $(\widetilde{x_i},\widetilde{y_i})$ since it is free from poisoning. While handling the sample with the target label in $P$, we utilize the consensus label in $C_{(x_i,y_i)}$ ($y_g$ in Equation~\ref{eq3}) as $\widetilde{y_i}$, and select $\widetilde{x_i}$ from $C_{(x_i,y_i)}$ based on the following insight: the forward and reverse process can effectively eliminate the trigger features while preserving the benign features of $x_i$. If the image $x_j$ from $C_{(x_i,y_i)}$ significantly differs from the poisoned image $x_i$, it is likely due to the effective removal of the trigger features on $x_j$. Therefore, we select $\widetilde{x_i}$ from $C_{(x_i,y_i)}$ by choosing $x_j$ that is distinct from $x_i$, thus only containing the benign features of $x_i$:
\begin{equation}  
\label{eq5}
\begin{aligned}
    dist(x_i,x_j) &= ssim(x_i,x_j) + cos(M(x_i),M(x_j)) \\
    &where ~x_j \in C_{(x_i,y_i)}
\end{aligned}
\end{equation}

To quantify the distance between $x_i$ and $x_j$, we employ \textit{Structure Similarity Index Measure}~\cite{wang2004image} in the pixel space and \textit{Cosine Similarity} in the feature space, using the victim model $M$ trained on the raw dataset $D$ as the feature extractor. We select $\widetilde{x_i}$ from $C_{(x_i,y_i)}$ that ranks at 80\% concerning the distance from $x_i$, to ensure the effective and comprehensive removal of trigger features.

For the sample $(x_i,y_i)$ with the target label in $S$, we opt to train a benign model $\widetilde{M}$ on the currently available purified dataset $\widetilde{D}$ to assist in determining its ground-truth label $\widetilde{y_i}$. The current $\widetilde{D}$ contains all the samples with benign labels, as well as those with the target label in $B \cup P$. The training of $\widetilde{M}$ involves two stages. In the first stage, we train $\widetilde{M}$ on $\widetilde{D}$ to predict the label $y_m$ for the suspicious sample $(x_i,y_i)$. Since $\widetilde{D}$ exclusively comprises either benign or purified samples, during training, $\widetilde{M}$ only learns benign features devoid of any backdoors, thus can assign the ground-truth labels to suspicious samples. Furthermore, we employ a voting mechanism based on the candidate set for the sample to obtain the label $y_v$, which is the most frequently occurring label in the candidate set. For the sample where $y_m$ is equal to $y_v$, we employ the resulting label as $\widetilde{y_i}$ and use Equation~\ref{eq5} to select $\widetilde{x_i}$ from the candidate set. Such purified sample $(\widetilde{x_i},\widetilde{y_i})$ is then added to $\widetilde{D}$ for the second stage of training, where we fine-tune $\widetilde{M}$ on the updated $\widetilde{D}$ with a lower learning rate to obtain the final benign model. Then each remaining suspicious sample is purified using $\widetilde{M}$, resulting in the final purified dataset $\widetilde{D}$.

\section{Evaluation}
\label{sec:evaluation}

\subsection{Experimental Setup} 
\label{subsec:experimental setup}

\noindent\textbf{Datasets and Models.} We conduct experiments on three datasets: CIFAR10~\cite{krizhevsky2009learning}, Tiny ImageNet~\cite{le2015tiny} and LFW~\cite{huang2008labeled}, using PreAct-ResNet18~\cite{he2016identity}, EfficientNet-B4~\cite{tan2019efficientnet}, VGGFace~\cite{parkhi2015deep} as classifier and DDPM~\cite{ho2020denoising} trained on CIFAR10, Improved DDPM~\cite{nichol2021improved} trained on ImageNet-1k~\cite{deng2009imagenet}, DDPM trained on CelebA-HQ~\cite{karras2018progressive} for purification.

\noindent\textbf{Attack Methods.} We evaluate nine popular attack methods, including BadNets~\cite{gu2019badnets}, Blended~\cite{chen2017targeted}, SSBA~\cite{li2020invisible}, 
LC~\cite{shafahi2018poison}, LF~\cite{zeng2021rethinking}, Ftrojan~\cite{wang2022invisible}, WaNet~\cite{nguyen2021wanet}, LIRA~\cite{doan2021lira} and IA~\cite{nguyen2020input}, using BackdoorBench~\cite{wu2022backdoorbench} for implementation.

\noindent\textbf{Parameter Settings.} The poison rate is set to 10\% and the target label is randomly selected without any specific strategy or preference. While using pre-trained diffusion models, we adhere to the default hyper-parameters provided by the source. The hyper-parameters of \tool~are set to: $T = 150$, $n = 5$, $m = 10$ and $\tau = 5$ ($0.1 \times n \times m$). We conduct preliminary experiments to study the impacts of hyper-parameters on CIFAR10 (Section~\ref{subsec:ablation study}) and determine these default settings, which also proved effective in countering diverse attacks across various datasets. 

\noindent\textbf{Metrics.} The metrics for evaluation are as follows:
\begin{itemize}
    \item \textit{True Positive Rate (TPR)} measures the probability that \tool~correctly detects poisoned samples in the raw dataset. Specifically, a sample is considered to be poisoned if its label in the raw dataset $D$ deviates from that in the purified dataset $\widetilde{D}$.
    \item \textit{False Positive Rate (FPR)} measures the probability that \tool~falsely detects benign samples as poisoned.
    \item \textit{Accuracy (ACC)} measures the probability that the model correctly classifies benign inputs.
    \item \textit{Attack Success Rate (ASR)} measures the probability that the model classifies poisoned inputs as the target label.
\end{itemize}

\subsection{Defense Performance}
\label{subsec:defense performance}

\begin{table*}[ht]
    \centering
    \footnotesize
    \tabcolsep=0.072cm
    \renewcommand\arraystretch{1.37}
    \scalebox{0.968}{\begin{tabular}{ccccccccccccccccccccc}
    \hline
    \multirow{2}{*}{\textbf{Attack}} &\multicolumn{4}{c}{\textbf{AC}} & \multicolumn{4}{c}{\textbf{Spectral}} &\multicolumn{4}{c}{\textbf{ABL}} & \multicolumn{4}{c}{\textbf{DBD}} & \multicolumn{4}{c}{\textbf{\tool}}  \\ \cmidrule(lr){2-5}\cmidrule(lr){6-9}\cmidrule(lr){10-13}\cmidrule(lr){14-17}\cmidrule(lr){18-21}
    \text{}& \textbf{TPR} & \textbf{FPR} & \textbf{ACC} & \textbf{ASR}
     & \textbf{TPR} & \textbf{FPR} & \textbf{ACC} & \textbf{ASR} & \textbf{TPR} & \textbf{FPR} & \textbf{ACC} & \textbf{ASR} & \textbf{TPR} & \textbf{FPR} & \textbf{ACC} & \textbf{ASR}
     & \textbf{TPR} & \textbf{FPR} & \textbf{ACC} & \textbf{ASR}\\
    \hline 
    \textbf{BadNets} & 	48.66	& 	48.81	& 	88.32	& 	99.61	& 	80.28	& 	4.41	& 	92.30	& 	98.00	& 	98.56	& \textbf{0.16} & 	86.90	& 	\textbf{0.00}	& 	0.22	& 	11.09	& 	76.55	& 	4.62	& 	\textbf{99.40}	& 	0.89 & 	\textbf{93.24}	& 	0.73\\ 
     \textbf{Blended}  &	48.06 &	48.84 &	88.86 &	95.62 &	98.04 &	2.44 &	92.66 &	57.44 &	86.64 &	1.48 &	85.56 &	\textbf{0.11} &	67.60 &	3.60 &	75.50 &	98.60  &	\textbf{99.18} &	\textbf{1.15} & \textbf{93.15} &	1.11\\
    \textbf{SSBA} &	48.50 &	48.62 &	88.37 &	94.41 &	54.38 &	7.29 &	90.74 &	87.16 &	57.72 &	4.70 &	87.37 &	7.84 &	50.76 &	5.47 &	75.99 &	4.55  &	\textbf{98.62} &	\textbf{0.87} & \textbf{93.23} &	\textbf{1.31} \\
    \textbf{LF} &	49.38 &	49.03 &	88.81 &	98.21 &	24.46 &	10.62 &	86.22 &	99.01 &	49.76 &	5.58 &	88.23 &	\textbf{0.96} &	50.96 &	5.45 &	75.49 &	5.26  &	\textbf{95.74} &	\textbf{0.96} & \textbf{93.65} &	4.61 \\
    \textbf{LC} &	48.20 &	49.12 &	89.39 &	11.40 &	85.20 &	8.23 &	90.29 &	2.34 &	0.00 &	1.01 &	82.00 &	6.00 &	0.00 &	1.01 &	76.57 &	5.45 & \textbf{92.80} & \textbf{3.82} & \textbf{93.37} & \textbf{1.91} \\
    \textbf{WaNet} &	48.89 &	48.52 &	89.14 &	91.79 &	47.65 &	7.90 &	91.06 &	87.59 &	20.29 &	8.94 &	89.09 &	82.62 &	52.43 &	5.61 &	76.36 &	5.94  &	\textbf{99.40} &	\textbf{1.20} & \textbf{92.61} &	\textbf{1.00} \\
    \textbf{IA} &	48.06 &	48.93 &	89.67 &	99.90 &	65.19 &	6.08 &	92.32 &	98.77 &	85.24 &	2.22 &	86.45 &	\textbf{0.18} &	52.50 &	5.60 &	75.22 &	6.39 &	\textbf{99.38} &	\textbf{0.79} &	\textbf{93.84} &	0.58 \\
    \textbf{LIRA} &	48.29 &	47.93 &	89.35 &	99.59 &	31.98 &	8.92 &	88.74 &	99.76 &	77.97 &	2.97 &	87.76 &	\textbf{0.24} &	54.74 &	5.37 &	76.94 &	3.69 & 	\textbf{98.81} &	\textbf{0.45} &	\textbf{93.68}	& 0.80 \\
    \cline{1-21}
     \textbf{Average} &	48.51 &	48.73 &	88.99 &	86.32 &	60.90 &	6.99 &	90.54 &	78.76 &	59.52 &	3.38 &	86.67 &	12.24 &	41.15 &	5.40 &	76.08 &	16.81 & \textbf{97.92} &	\textbf{1.27} &	\textbf{93.35} &	\textbf{1.51} \\
    \hline
   \textbf{BadNets} &	35.44 &	35.86 &	62.64 &	100.00 &	94.50 &	0.00 &	65.39 &	99.46 &	96.77 &	0.36 &	62.17 &	0.00 &	95.24 &	0.53 &	36.65 &	99.97 & \textbf{99.96} & \textbf{0.15} & \textbf{65.64} & 0.05 \\
    \textbf{Blended} &	36.45 &	36.71 &	61.84 &	95.21 &	94.14 &	\textbf{0.04} &	\textbf{66.07} &	7.83 &	79.93 &	2.23 &	59.58 &	\textbf{0.12} &	95.25 &	0.53 &	33.14 &	99.14 &	\textbf{97.52} &	0.16 &	65.77 &	0.48 \\
    \textbf{SSBA} &	33.94 &	34.46 &	62.71 &	98.18 &	89.93 &	0.51 &	65.52 &	91.30 &	85.55 &	1.61 &	59.35 &	\textbf{0.07} &	95.25 &	0.53 &	10.09 &	98.05 & \textbf{99.66} & \textbf{0.16} & \textbf{66.69} & 0.41 \\
    \textbf{LF} &	34.68 &	35.10 &	62.53 &	96.23 &	90.41 &	0.45 &	65.45 &	56.49 &	61.84 &	4.24 &	59.11 &	64.47 &	95.20 &	0.53 &	26.03 &	93.01 &	\textbf{95.96} &	\textbf{0.14} &	\textbf{65.74} &	\textbf{2.08} \\
    \textbf{Ftrojan} &	35.38 &	35.62 &	61.72 &	99.91 &	94.45 &	\textbf{0.00} &	65.67 &	13.42 &	74.58 &	2.82 &	61.55 &	\textbf{0.01} &	95.28 &	0.52 &	29.65 &	99.75 & \textbf{98.11} & 0.13 & \textbf{66.12} & 1.85 \\
    \textbf{WaNet} &	36.25 &	35.73 &	61.73 &	95.93 &	94.10 &	\textbf{0.05} &	64.96 &	5.84 &	34.69 &	7.33 &	62.46 &	\textbf{0.00} &	97.39 &	0.54 &	32.80 &	98.56 &	\textbf{99.72} &	0.14 & \textbf{65.18} &	0.16 \\
    \textbf{IA} &	37.15 &	36.79 &	58.55 &	99.89 &	90.92 &	3.99 &	65.65 &	97.98 &	87.72 &	1.59 &	58.90 &	\textbf{0.01} &	97.41 &	0.54 &	33.48 &	99.94 &	\textbf{95.00} &	\textbf{0.18} &	\textbf{64.96} &	0.51 \\
    \cline{1-21}
     \textbf{Average} & 35.61 & 35.75 &61.67 & 97.91 & 92.64 & 0.72 & 65.53 & 53.19 & 74.44 &	2.88 &	60.45 &	9.24 &	95.86 &	0.53 &	28.83 &	98.35 &	\textbf{97.71} &	\textbf{0.15} &	\textbf{65.73} &	\textbf{0.79}\\
    \hline
\end{tabular} }
\caption{Defense performance on CIFAR10 (upper part) and Tiny ImageNet (lower part).}
\label{tab1}
\end{table*}

As shown in Table~\ref{tab1}, we compare \tool~with four baseline defense methods aimed at dataset sanitation, including AC~\cite{chen2019detecting}, Spectral~\cite{tran2018spectral}, ABL~\cite{li2021anti} and DBD~\cite{huang2022backdoor}. Notably, \tool~effectively defends against all nine evaluated backdoor attacks, achieving the highest TPR, the lowest FPR, the highest ACC, and the lowest ASR, with averages of 97.92\%, 1.27\%, 93.35\%, 1.51\% on CIFAR10 and 97.71\%, 0.15\%, 65.73\% and 0.79\% on Tiny ImageNet. In contrast, the efficacy of baseline methods varies significantly across different attacks and datasets. Specifically, AC exhibits limited performance against diverse attacks, Spectral is effective only for certain attacks, and DBD fails on Tiny ImageNet. While ABL shows lower ASR against some attacks, it does so by sacrificing benign accuracy and also proves inadequate against WaNet on CIFAR10 and LF on Tiny ImageNet. In comparison, \tool~exhibits excellent defense effectiveness while maintaining the model's performance on the benign task, with an average increase of 0.26\% on CIFAR10 and 0.91\% on Tiny ImageNet compared to the victim model.

\begin{table}[ht]
\centering
\footnotesize
\tabcolsep=0.09cm
\renewcommand\arraystretch{1.5}
\begin{tabular}{ccccccc}
\hline
\multirow{2}{*}{\textbf{Attack}} &\multicolumn{2}{c}{\textbf{No Defense}} & \multicolumn{4}{c}{\textbf{\tool}} \\ \cmidrule(lr){2-3}\cmidrule(lr){4-7}
\text{}& \textbf{ACC} & \textbf{ASR} & \textbf{TPR} & \textbf{FPR} & \textbf{ACC} & \textbf{ASR}\\
\hline 
\textbf{BadNets} & 94.52 & 99.84 & 98.33 &	1.34 & 94.03 & 0.82 \\ 
\textbf{Blended} & 94.68 & 99.34 & 99.58 & 1.25 & 94.35 & 0.16 \\
\textbf{LF} & 92.90 & 94.10 & 97.92 & 1.34 & 94.19 & 0.49 \\
\textbf{Ftrojan} & 94.52 & 93.44 & 99.17 &	1.34 & 94.35 & 0.00 \\
\hline
\end{tabular}
\caption{Defense performance on LFW.}
\label{tab2}
\end{table}

Moreover, we compare \tool~with other types of defense methods including model reconstruction~\cite{wang2019neural,li2021neural,wu2021adversarial,zeng2022adversarial} and input purification~\cite{doan2020februus,shi2023blackbox,may2023salient}. Our approach exhibits superior defense performance. We also validate the effectiveness of \tool~on LFW, and the results are shown in Table~\ref{tab2}. \tool~effectively mitigates all attacks, achieving the average TPR, FPR, ACC and ASR of 98.75\%, 1.32\%, 94.23\% and 0.37\%, respectively. 

In addition, \tool~exhibits excellent performance in target label detection. While countering poisoned CIFAR10, \tool~accurately detects the target label with an average anomaly index of 5.4323, and produces no false positives for benign labels with the average maximum anomaly index of 1.0010. For benign CIFAR10, \tool~also avoids false positives, with the maximum anomaly index of 1.6992, below the threshold value of 2.

\subsection{Evaluation on Purified Dataset}
\label{subsec:evaluation on purified dataset}

We conduct experiments to evaluate the purified datasets generated by \tool~in terms of label accuracy, image quality, and the performance of newly trained models. The average label accuracy is 98.10\% on CIFAR10 and 95.00\% on Tiny ImageNet. Besides, the average \textit{Fréchet Inception Distance (FID)}~\cite{heusel2017gans} and \textit{Inception Score (IS)}~\cite{salimans2016improved} of purified images are 0.71 and 9.99 on CIFAR10, and 0.35 and 28.62 on Tiny ImageNet. The newly trained models on these purified datasets achieve an average ACC and ASR of 93.05\% and 1.13\% on CIFAR10 and 65.75\% and 0.41\% on Tiny ImageNet.

\subsection{Defense against Various Attack Scenarios}
\label{subsec:defense against various attack scenarios}

We evaluate the performance of \tool~against various attack scenarios using datasets poisoned with poison rates in $\{1\%, 3\%, 5\%, 7\%, 10\%\}$, target label numbers in $\{1, 2, 3\}$ and target label selected in $\{0, 2, 4, 6, 8\}$. In all scenarios, our approach demonstrates effective performance, with the average TPR, FPR, ACC, ASR of 99.00\%, 0.65\%, 93.68\%, 0.62\% for varying poison rates, 99.10\%, 1.20\%, 93.34\%, 0.54\% for varying target label numbers, and 99.16\%, 0.82\%, 93.47\%, 0.71\% for different target labels, validating the robustness of \tool.

\subsection{Ablation Study}
\label{subsec:ablation study}

\noindent\textbf{Impact of Diffusion Models.} We evaluate the impact of diffusion models representing different distributions on CIFAR10. In addition to the identical distribution using DDPM trained on CIFAR10, we consider the similar distribution using Improved DDPM trained on ImageNet-1k and disparate distribution using DDPM trained on CelebA~\cite{liu2015deep}. As shown in Table~\ref{tab3}, the similar diffusion model achieves results comparable to those of the identical diffusion model. On the other hand, the disparate diffusion model also proves effective in countering various attacks, while it slightly affects the benign accuracy due to its partial restoration of degraded benign features. However, it still surpasses the performance of the baseline defense methods outlined in Table~\ref{tab1}. These results validate the practical applicability of \tool~in realistic scenarios.

\begin{table}[ht]
\centering
\footnotesize
\tabcolsep=0.09cm
\renewcommand\arraystretch{1.5}
\begin{tabular}{ccccccccc}
\hline
\multirow{2}{*}{\textbf{Attack}} &\multicolumn{4}{c}{\textbf{Similar Distribution}} & \multicolumn{4}{c}{\textbf{Disparate Distribution}} \\ \cmidrule(lr){2-5}\cmidrule(lr){6-9}
\text{}& \textbf{TPR} & \textbf{FPR} & \textbf{ACC} & \textbf{ASR} & \textbf{TPR} & \textbf{FPR} & \textbf{ACC} & \textbf{ASR}\\
\hline 
\textbf{BadNets} & 98.54 & 1.56 & 92.47 & 0.82 & 97.84 & 4.62 & 89.53 & 1.36 \\ 
\textbf{Blended} & 98.72 & 1.73 & 92.52 & 0.70 & 98.52 & 4.69 & 90.45 & 1.20 \\
\textbf{SSBA} & 99.10 & 1.44 & 92.82 & 0.59 & 99.00 & 6.79 & 89.34 & 0.76 \\
\textbf{IA} & 99.17 & 1.34 & 92.65 & 0.56 & 98.44 & 4.17 & 90.50 & 0.66 \\ 
\textbf{LIRA} & 99.02 & 1.19 & 93.12 & 0.69 & 98.46 & 3.49 & 91.28 & 0.64 \\
\hline
\end{tabular}
\caption{Impact of distribution of diffusion model.}
\label{tab3}
\end{table}

\begin{figure}[htbp]
  \centering
  \begin{subfigure}[b]{0.42\linewidth}
        \centering
        \includegraphics[width=\columnwidth]{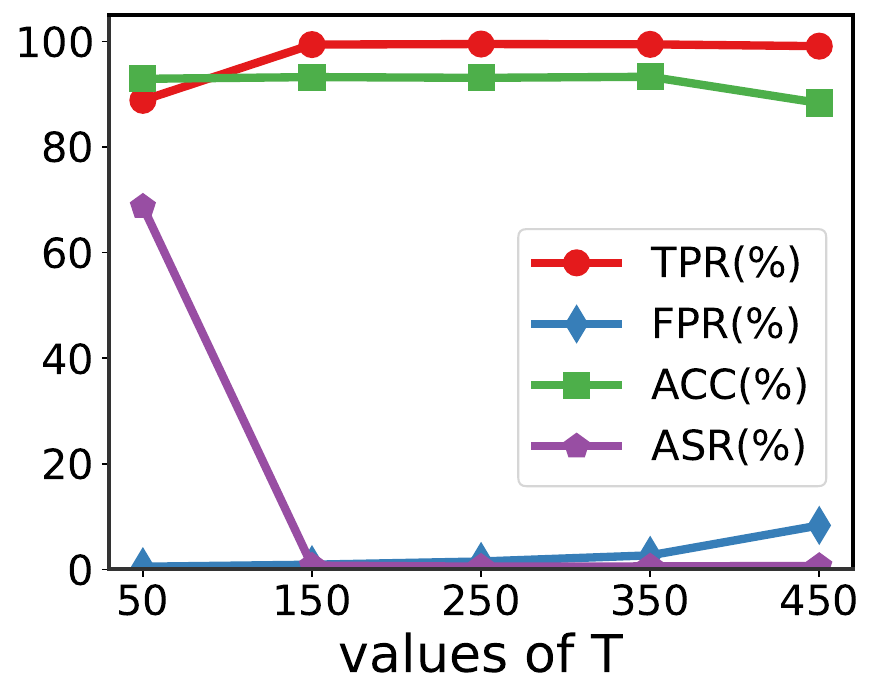}
        \caption{Impact of $T$}
        \label{fig3-1}
  \end{subfigure}
  \hspace{0.8cm}
  \begin{subfigure}[b]{0.42\linewidth}
        \centering
        \includegraphics[width=\columnwidth]{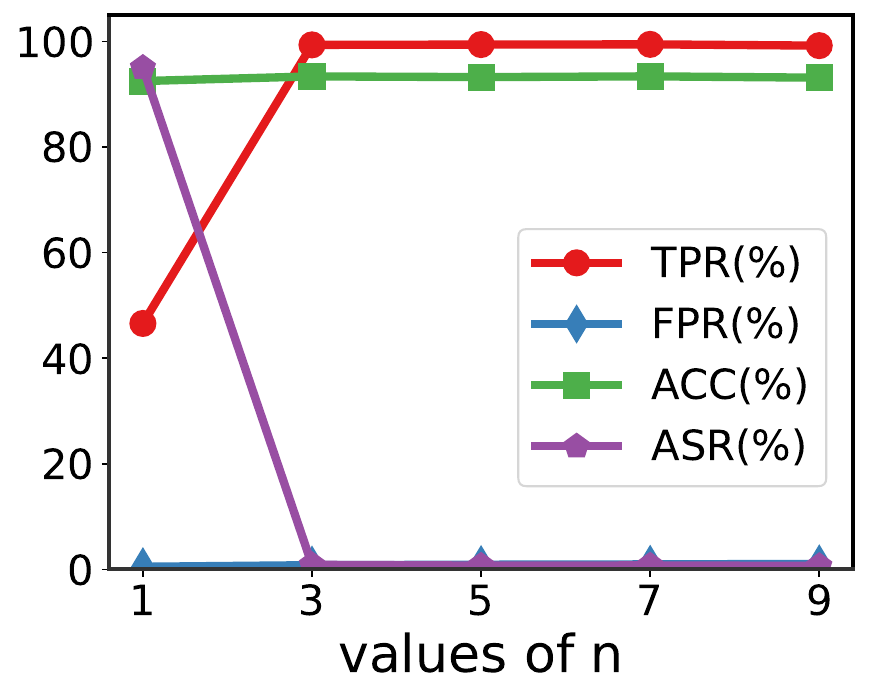}
        \caption{Impact of $n$}
        \label{fig3-2}
  \end{subfigure}
  \hfill
  \begin{subfigure}[b]{0.42\linewidth}
        \centering
        \includegraphics[width=\columnwidth]{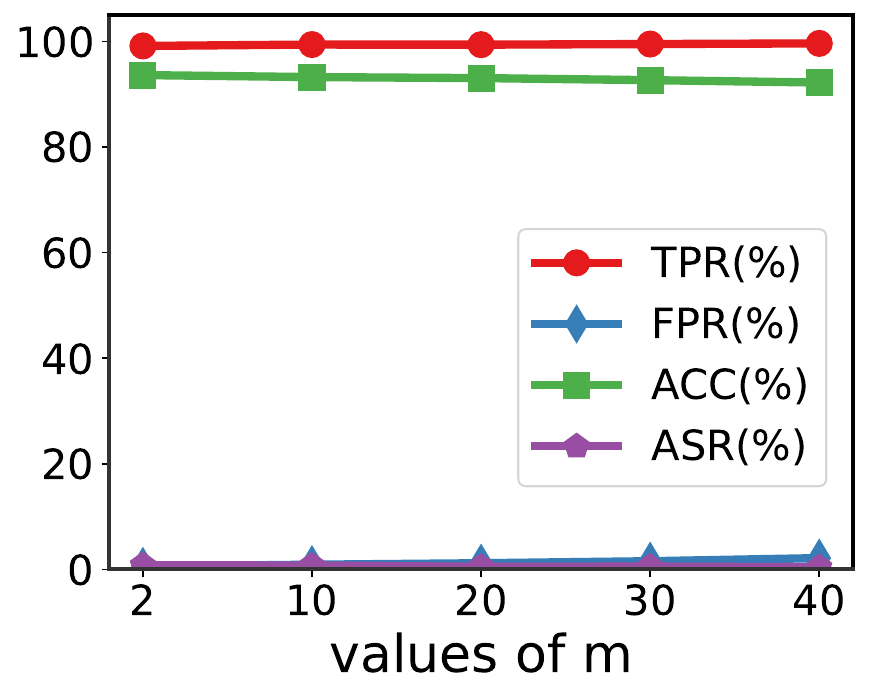}
        \caption{Impact of $m$}
        \label{fig3-3}
  \end{subfigure}
  \hspace{0.8cm}
  \begin{subfigure}[b]{0.42\linewidth}
        \centering
        \includegraphics[width=\columnwidth]{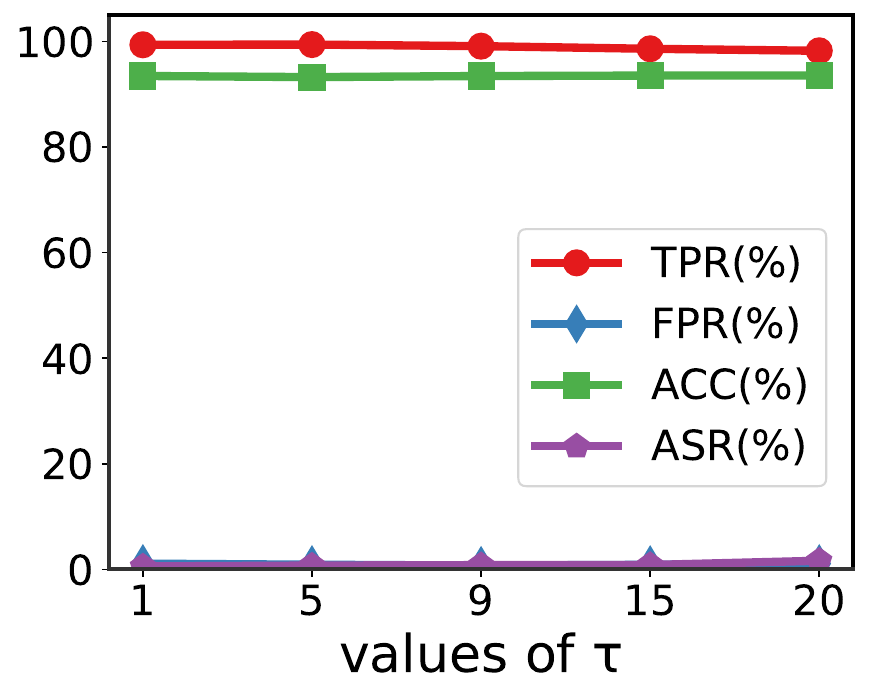}
        \caption{Impact of $\tau$}
        \label{fig3-4}
  \end{subfigure}
  \caption{Impact of hyper-parameters.}
  \label{fig3}
\end{figure}

\noindent\textbf{Impact of Hyper-parameters.} As shown in Figure~\ref{fig3}, we assess the impact of hyper-parameters on the performance of \tool. Our experiments are conducted on CIFAR10 using BadNets. $T$ and $n$ mainly influence the magnitude of the noise introduction, where lower values lead to lower TPR and higher ASR, indicating ineffective removal of trigger features, and higher values of $T$ lead to higher FPR and lower ACC, indicating excessive degradation of benign features. $m$ dictates when we extract intermediary images from the reverse process, and higher values correspond to higher FPR and lower ACC, indicating incomplete restoration of benign features. $\tau$ determines the strictness of label transition judgment, with higher values signifying more lenient judgments and resulting in lower TPR and higher ASR.  

We also evaluate the impacts of our proposed purified image selection strategy and ground-truth label determination strategy, which lead to the increase of 58.88\% and 9.92\% in TPR, along with the decrease of 97.71\% and 62.90\% in ASR compared to the baseline strategy.

\section{Discussion}
\label{sec:discussion}

\subsection{Adaptive Attack}
\label{subsec:adaptive attack}

\noindent\textbf{Poisoned Diffusion Model.} We validate the performance of \tool~when using diffusion model trained on the poisoned dataset. Our experiments are conducted on CIFAR10, using the diffusion model trained on data poisoned by BadNets. As shown in Table~\ref{tab4}, this poisoned diffusion model fails to counter the corresponding backdoor attack. Nevertheless, it retains the ability to purify training datasets poisoned by various other backdoor attacks, all without introducing BadNets into the resulting purified datasets. Consequently, in scenarios where access to trustworthy data and diffusion models are unavailable, we propose training the diffusion model using data from diverse untrustworthy sources. Notably, \tool~performs effectively as long as the data used for diffusion model training and the data intended for purification are not poisoned by the same trigger.

\begin{table}[ht]
\centering
\footnotesize
\tabcolsep=0.09cm
\renewcommand\arraystretch{1.5}
\begin{tabular}{ccccccccc}
    \hline
    \textbf{Attack}	& \textbf{BadNets} & \textbf{Blended} & \textbf{SSBA} & \textbf{IA} & \textbf{LIRA}\\
    \hline
    \textbf{TPR} & 85.48 & 98.52 & 99.02 & 99.57 & 98.87\\
    \textbf{FPR} & 1.00 & 1.41 & 1.34 & 1.08 & 0.58\\
    \textbf{ACC} & 91.66 & 93.02 & 93.09 & 93.17 & 93.61\\
    \textbf{ASR} & 99.76 & 1.54 & 0.97 & 0.71 & 0.78\\
    \textbf{ASR-BadNets} & 99.76 & 0.64 & 0.50 & 0.62 & 0.71\\
    \hline	
\end{tabular}
\caption{Defense performance using diffusion model trained on poisoned dataset.}
\label{tab4}
\end{table}

\noindent\textbf{Residual Backdoor.} We assume that the attacker possesses comprehensive knowledge of the defense mechanism of \tool. The attacker considers an adaptive attack presuming that residual trigger features remain in purified images, and such remaining features can inject a residual backdoor into the model trained on them (different from the original backdoor injected by the full trigger features). Specifically, the attacker employs the same purification strategy to generate poisoned inputs that contain the same residual trigger features, aiming to activate the residual backdoor in the benign model trained on the purified dataset. We evaluate such adaptive attack on CIFAR10, and the results are shown in Table~\ref{tab5}. Although the defense results are slightly higher than those outlined in Section~\ref{subsec:defense performance}, all ASRs remain below 5\%, indicating that such residual backdoor is ineffective.

\begin{table}[ht]
\centering
\footnotesize
\tabcolsep=0.09cm
\renewcommand\arraystretch{1.5}
\begin{tabular}{ccccccccc}
    \hline
    \textbf{Attack}	& \textbf{BadNets} & \textbf{Blended} & \textbf{SSBA} & \textbf{IA} & \textbf{LIRA}\\
    \hline
    \textbf{ASR} & 3.12 & 3.77 & 4.46 & 4.06 & 4.10\\
    \hline	
\end{tabular}
\caption{ASR of the potential residual backdoor.}
\label{tab5}
\end{table}

\subsection{Limitation}
\label{subsec:limitation}

Efficient image sampling has been a persistent challenge for diffusion models since their inception, which also limits our approach, as Candidate Set Construction constitutes approximately 65\% of the overall defense duration. However, since the process of constructing candidate set is sample-wise, parallelization techniques can be employed to enhance efficiency while additional computational resources are available. In addition, several methods~\cite{song2021denoising,lu2022dpm} aimed at accelerating the sampling efficiency of diffusion models have been proposed, we plan to incorporate these methods in \tool~in future work.

\section{Conclusion}
\label{sec:conclusion}

In this paper, we propose \tool, a novel dataset sanitization approach to purify poisoned training datasets using diffusion models to effectively defend against poisoning-based backdoor attacks. Our method utilizes the forward and reverse process to construct the candidate set for each sample, enabling the identification of anomalous samples, detection of target labels, selection of purified images, and determination of their ground-truth labels. Experimental results validate that \tool~can effectively mitigate diverse backdoor attacks while preserving the benign accuracy, outperforming existing backdoor defense methods.

\section*{Acknowledgments}
We thank all the anonymous reviewers for their constructive feedback. This work is supported in part by National Key Research and Development Program (2020AAA0107800), NSFC (62302498, 92270204), Youth Innovation Promotion Association CAS, Beijing Nova Program and a research grant from Huawei.

\bibliography{aaai24}

\appendix
\begin{center}
   \LARGE \textbf{Appendix} 
\end{center}

\section{Details of Experimental Setup}
\label{sec:details of experimental setup}

\subsection{Details of Datasets and Models}
\label{subsec:details of datasets and models}

\begin{itemize}
    \item \textbf{CIFAR10}~\cite{krizhevsky2009learning} is a widely used benchmark dataset in computer vision that consists of 10 different types of objects and animals. Each image has a resolution of $32 \times 32$ pixels and is represented in RGB format. We train the PreAct-ResNet18~\cite{he2016identity} model from scratch as classifier, and use the pre-trained DDPM~\cite{ho2020denoising} for purification.
    \item \textbf{Tiny ImageNet}~\cite{le2015tiny} is a subset of the well-known object classification dataset ImageNet~\cite{deng2009imagenet} with 200 object classes. Each image has a size of $64 \times 64$ pixels in RGB format. We fine-tune the EfficientNet-B4~\cite{tan2019efficientnet} model pre-trained on ImageNet-1k as classifier, and use the Improved DDPM~\cite{nichol2021improved} trained on ImageNet-1k for purification. 
    \item \textbf{LFW}~\cite{huang2008labeled} is a popular human face recognition dataset. We collect classes with more than 20 images, resulting in 62 human face classes. These images are resized to $224 \times 224$ pixels. We fine-tune the pre-trained VGGFace~\cite{parkhi2015deep} model as classifier, and use the DDPM trained on CelebA-HQ~\cite{karras2018progressive} for purification.
\end{itemize}

\subsection{Details of Backdoor Attacks}
\label{subsec:details of backdoor attacks}

\begin{itemize}
    \item \textbf{BadNets}~\cite{gu2019badnets} is the pioneering backdoor attack, which creates poisoned samples by stamping patterns (\eg, 3 × 3 white square) on the corners of benign images and modifying their labels to the target label. 
    \item \textbf{Blended}~\cite{chen2017targeted} generates poisoned images by blending benign images with the key pattern (cartoon images or random patterns). 
    \item \textbf{SSBA (Sample Specific Backdoor Attack)}~\cite{li2020invisible} utilizes an auto-encoder to incorporate the trigger (\eg, a string) into benign images, resulting in unique trigger specific to each benign image. 
    \item \textbf{LF (Low Frequency)}~\cite{zeng2021rethinking} proposes an optimization-based approach to create smooth triggers, effectively reducing the high-frequency artifacts. 
    \item \textbf{Ftrojan}~\cite{wang2022invisible} poisons benign images by applying low-frequency perturbations. 
    \item \textbf{WaNet (Warping-based poisoned Network)}~\cite{nguyen2021wanet} uses warping functions to slightly distort benign images, thereby constructing poisoned images. 
    \item \textbf{IA (Input-Aware)}~\cite{nguyen2020input} trains a trigger generator simultaneously with the backdoor model, capable of generating various triggers for benign images. 
    \item \textbf{LIRA (Learnable, Imperceptible and Robust backdoor Attack)}~\cite{doan2021lira} formulates backdoor injection as a non-convex, constrained optimization problem and trains a generator as the poisoning function. 
    \item \textbf{LC (Label Consistent)}~\cite{shafahi2018poison} is a clean-label attack that does not require modifying the labels of poisoned samples to the target label. Instead, it injects triggers using adversarial attack methods to generate poisoned images that resemble both benign target images and the original images.
\end{itemize} 

\subsection{Details of Baseline Backdoor defenses}
\label{subsec:details of baseline backdoor defenses}
  
\begin{itemize}
    \item \textbf{AC (Activation Clustering)}~\cite{chen2019detecting} analyzes the neural network activations of training samples to determine whether they are poisoned.
    \item \textbf{Spectral}~\cite{tran2018spectral} reveals that poisoned samples typically produce a detectable trace in the spectrum of the covariance of feature representations.  
    \item \textbf{ABL (Anti-Backdoor Learning)}~\cite{li2021anti} utilizes a two-stage gradient ascent mechanism for training, isolating backdoor examples at an early training stage and breaking the correlation between backdoor examples and the target label at a later training stage.
    \item \textbf{DBD (Decoupling-based Backdoor Defense)}~\cite{huang2022backdoor} first trains the backbone model via self-supervised learning and then trains the remaining fully-connected layers via supervised learning, followed by further removing the backdoor by conducting a semi-supervised fine-tuning of the entire model.
\end{itemize}

\section{Comparison with Other Defense Methods}
\label{sec:comparison with other defense methods}

\subsection{Details of Other Backdoor Defenses}
\label{subsec:details of other backdoor defenses}

\begin{table*}[ht]
\centering
\footnotesize
\tabcolsep=0.09cm
\renewcommand\arraystretch{1.5}
\begin{tabular}{cccccccccccccc}
\hline
\multirow{2}{*}{\textbf{Dataset}} & \multirow{2}{*}{\textbf{Attack}} &\multicolumn{2}{c}{\textbf{No Defense}} &\multicolumn{2}{c}{\textbf{NC}} & \multicolumn{2}{c}{\textbf{ANP}} &\multicolumn{2}{c}{\textbf{I-BAU}} & \multicolumn{2}{c}{\textbf{NAD}} & \multicolumn{2}{c}{\textbf{\tool}}\\ \cmidrule(lr){3-4}\cmidrule(lr){5-6}\cmidrule(lr){7-8}\cmidrule(lr){9-10}\cmidrule(lr){11-12} \cmidrule(lr){13-14}
\text{}&\text{}& \textbf{ACC} & \textbf{ASR} & \textbf{ACC} & \textbf{ASR} & \textbf{ACC} & \textbf{ASR} & \textbf{ACC} & \textbf{ASR} & \textbf{ACC} & \textbf{ASR} & \textbf{ACC} & \textbf{ASR}\\
\hline 
\multirow{9}{*}{\rotatebox{90}{\textbf{CIFAR10}}} & \textbf{BadNets}	& 92.39 & 99.96 &	91.44 &	1.09 &	90.33 &	1.92 &	89.77 &	0.94 &	90.68 &	3.17 &	\textbf{93.24} &	\textbf{0.73} \\ 
 & \textbf{Blended}	&  93.22 &	98.76 &	93.22 &	98.76 &	88.96 &	14.02 &	90.51 &	9.16 &	92.30 &	96.09 &	\textbf{93.15} &	\textbf{1.11} \\
& \textbf{SSBA} &  93.03 & 98.72 &	91.36 &	\textbf{0.59} &	90.06 &	0.96 &	90.96 &	4.00 &	92.28 &	80.74 &	\textbf{93.23} &	1.31 \\
& \textbf{LF} & 92.94 & 99.48 &	91.65 &	\textbf{1.91} &	88.95 &	5.63 &	90.08 &	3.21 &	92.40 &	94.21 &	\textbf{93.65} & 4.61	 \\
& \textbf{LC} & 93.85 & 98.98 &	92.48 &	4.37 &	92.96 &	3.67 &	91.96 &	\textbf{1.68} &	93.76 &	94.28 &	\textbf{93.37} & 1.91 \\
& \textbf{WaNet} & 92.80 & 96.20 &	91.33 &	1.24 &	88.62 &	6.42 &	89.39 &	\textbf{0.57} &	91.92 &	49.06 &	\textbf{92.61} &	1.00 \\
& \textbf{IA} & 92.80 &	86.87 &	92.80 &	86.87 &	88.21 &	0.86 &	88.57 &	3.99 &	91.81 &	39.53 &	\textbf{93.84} &	\textbf{0.58} \\
& \textbf{LIRA} &  93.67 & 99.92 &	91.72 &	0.94 &	91.57 &	\textbf{0.73} &	90.31 &	6.79 &	93.15 &	38.86 &	\textbf{93.68} &	0.80 \\
\cline{2-14}
 & \textbf{Average} & 93.09 & 97.36 &	92.00 &	24.47 &	89.96 &	4.28 &	90.19 &	3.79 &	92.29 &	61.99 &	\textbf{93.35} &	\textbf{1.51} \\
\hline
\multirow{9}{*}{\rotatebox{90}{\textbf{Tiny ImageNet}}} & \textbf{BadNets}	& 65.46 & 100.00 &	61.66 &	0.76 &	59.43 &	10.74 &	56.04 &	99.91 &	62.77 &	0.06 &	\textbf{65.64} &	\textbf{0.05} \\
& \textbf{Blended} &  64.42 & 97.35 &	58.90 &	0.08 &	61.47 &	\textbf{0.00} &	56.96 &	84.88 &	61.07 &	1.12 &	\textbf{65.77} &	0.48 \\
& \textbf{SSBA} & 65.16 & 98.11 &	61.62 &	2.51 &	57.57 &	79.28 &	57.83 &	92.88 & 62.99 & 14.59 & \textbf{66.69} & \textbf{0.41}\\
& \textbf{LF} & 62.17 & 96.33 &	57.11 &	\textbf{0.44} &	55.59 &	87.35 &	56.33 &	84.37 &	55.95 &	1.90 &	\textbf{65.74} &	2.08 \\
& \textbf{Ftrojan} & 65.74 & 99.98 &	61.69 &	0.28 &	64.51 &	38.83 &	56.31 &	98.31 &	62.91 &	\textbf{0.08} &	\textbf{66.12} &	1.85 \\
& \textbf{WaNet} & 65.09 & 97.63 &	61.15 &	2.03 &	62.61 &	1.63 &	55.48 &	75.12 &	63.27 &	3.38 &	\textbf{65.18} &	\textbf{0.16} \\
& \textbf{IA} & 65.72 & 93.42 &	63.36 &	1.76 &	65.68 &	93.37 &	57.15 &	42.32 &	62.91 &	10.68 &	\textbf{64.96} &	\textbf{0.51} \\
\cline{2-14}
 & \textbf{Average} & 64.82 & 97.55 & 60.78 & 1.12 & 60.98 & 44.46 & 56.69 & 82.54 & 61.70 & 4.54 & \textbf{65.73} & \textbf{0.79}\\
\hline
\end{tabular}
\caption{Comparison with model reconstruction methods.}
\label{tab6}
\end{table*}

In Section~\ref{subsec:defense performance}, we compare \tool~with four baseline defense methods aimed at dataset sanitization. In this section, we extend the comparison to other types of backdoor defense methods, including model reconstruction~\cite{wang2019neural,li2021neural,zeng2022adversarial,wu2021adversarial} and input purification~\cite{doan2020februus,shi2023blackbox,may2023salient}. 

\noindent\textbf{Model Reconstruction.} Model reconstruction aims to directly mitigate backdoors in the compromised model. 

\begin{itemize}
    \item \textbf{NC (Neural Cleanse)}~\cite{wang2019neural} reverses the potential trigger through optimization and then mitigates the backdoor by pruning the poisoned neurons or fine-tuning the model weights based on the reversed trigger. 
    \item \textbf{ANP (Adversarial Neuron Pruning)}~\cite{wu2021adversarial} focuses on eliminating backdoors by pruning the poisoned neurons in the backdoor model that are sensitive to adversarial neuron perturbation.
    \item \textbf{I-BAU (Implicit Backdoor Adversarial Unlearning)}~\cite{zeng2022adversarial} presents a min-max formulation to remove backdoors in the backdoor model, where the inner maximum seeks a powerful trigger that leads to a high loss, while the outer minimum aims to suppress this adversarial loss to unlearn the injected backdoor.
    \item \textbf{NAD (Neural Attention Distillation)}~\cite{li2021neural} initially fine-tunes the backdoor model as the teacher model and then performs another fine-tuning process by encouraging coherence in the intermediate-layer attention between the benign model and the teacher model.
\end{itemize} 

\noindent\textbf{Input Purification.} Input purification utilizes generative models to purify poisoned inputs. However, these methods primarily focus on preventing the activation of backdoors in the infected model rather than clearing them from existence, while \tool~aims at purifying poisoned datasets to train benign models once for all. 

\begin{itemize}
    \item \textbf{Februus}~\cite{doan2020februus} employs GradCAM to localize and remove the trigger from the poisoned input, then utilizes a GAN to restore the damaged regions.
    \item \textbf{Sancdifi (Salient conditional diffusion)}~\cite{may2023salient} identifies the trigger regions on the poisoned input using GradCAM, and removes the trigger and reconstructs the image using diffusion models.
    \item \textbf{ZIP (Zero-shot Image Purification)}~\cite{shi2023blackbox} applies a linear transformation to the poisoned input to eliminate the trigger, and then recovers the missing semantic information using a pre-trained diffusion model.
\end{itemize}

\subsection{Comparison with Model Reconstruction}
\label{subsec:comparison with model reconstruction methods}

We compare \tool~with four model reconstruction methods, and the evaluation results are shown in Table~\ref{tab6}. \tool~demonstrates superior defense performance, achieving the highest ACC and the lowest ASR, with averages of 93.35\%, 1.51\% on CIFAR10 and 65.73\% and 0.79\% on Tiny ImageNet, respectively. In contrast, the other methods exhibit limited effectiveness across diverse attacks and datasets. Specifically, ANP and I-BAU show ineffective defense on Tiny ImageNet while NAD performs poorly on CIFAR10. Moreover, while NC successfully counters most attacks (except for Blended and IA on CIFAR10), it negatively affects the ACC of the model, with an average decrease of 1.09\% on CIFAR10 and 4.04\% on Tiny ImageNet. Conversely, \tool~effectively mitigates all types of attacks without compromising the ACC, with an average increase of 0.26\% on CIFAR10 and 0.91\% on Tiny ImageNet.

\begin{figure*}[!t]
   \centering
    \begin{subfigure}[b]{0.24\linewidth}
        \centering
        \includegraphics[width=\columnwidth]{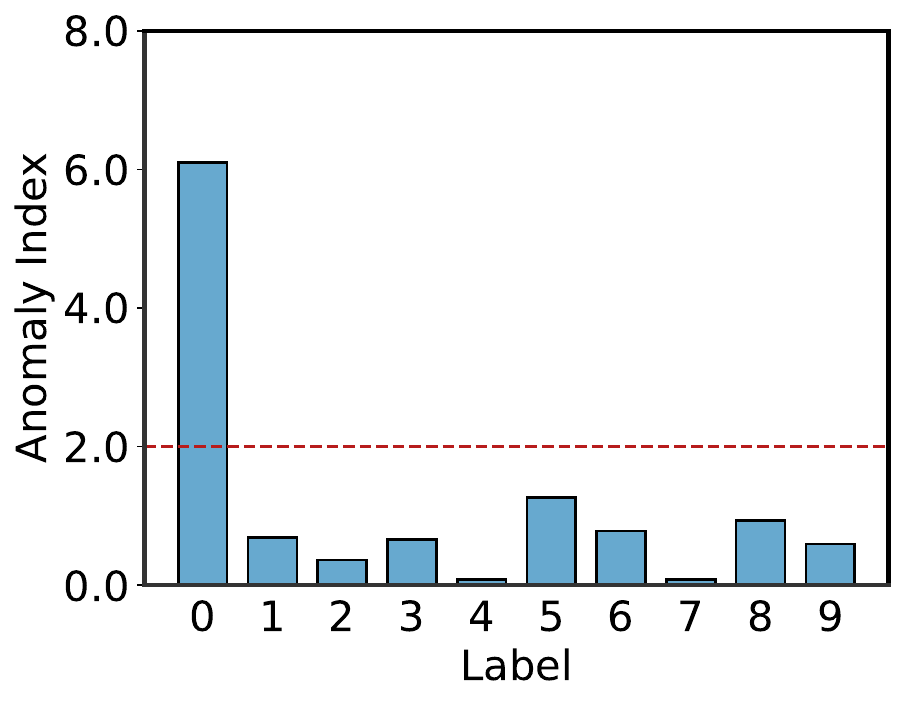}
        \caption{BadNets}
        \label{fig4-1}
    \end{subfigure}
  \hfill
  \begin{subfigure}[b]{0.24\linewidth}
        \centering
        \includegraphics[width=\columnwidth]{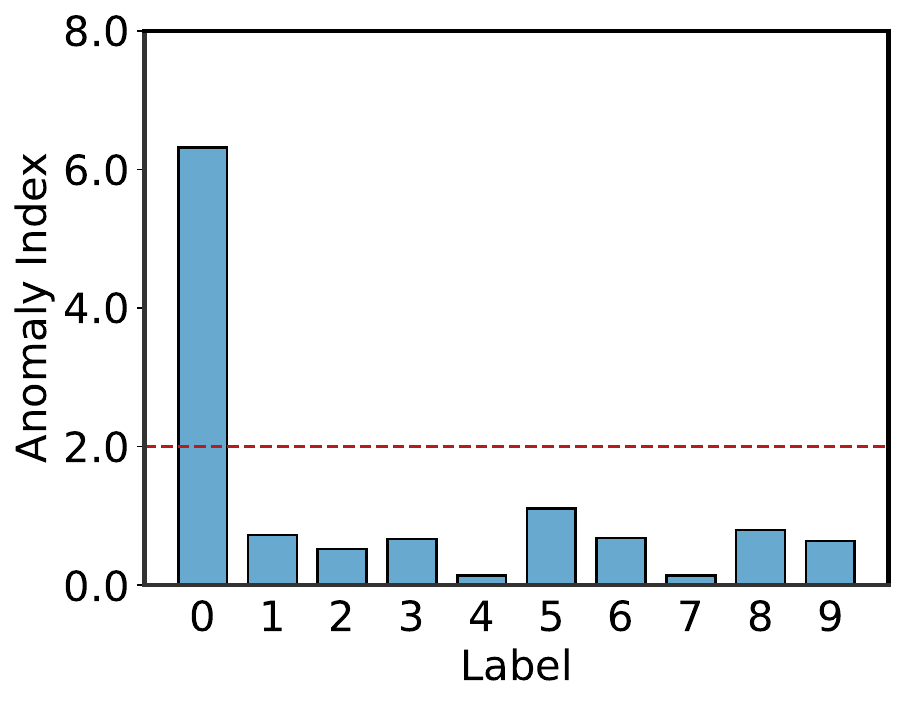}
        \caption{Blended}
        \label{fig4-2}
    \end{subfigure}
  \hfill
  \begin{subfigure}[b]{0.24\linewidth}
        \centering
        \includegraphics[width=\columnwidth]{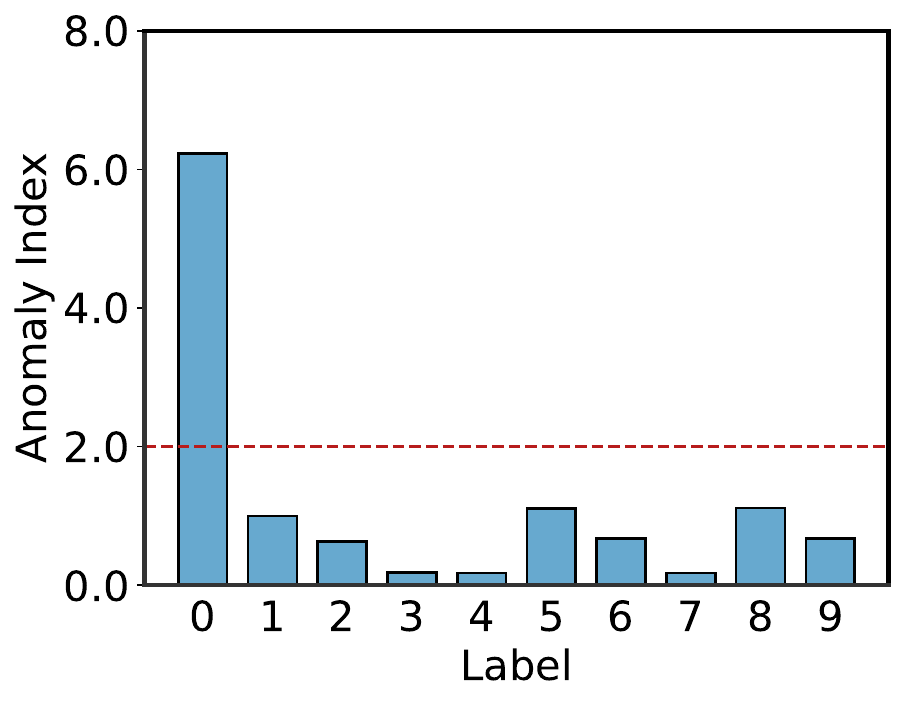}
        \caption{SSBA}
        \label{fig4-3}
    \end{subfigure}
  \hfill
  \begin{subfigure}[b]{0.24\linewidth}
        \centering
        \includegraphics[width=\columnwidth]{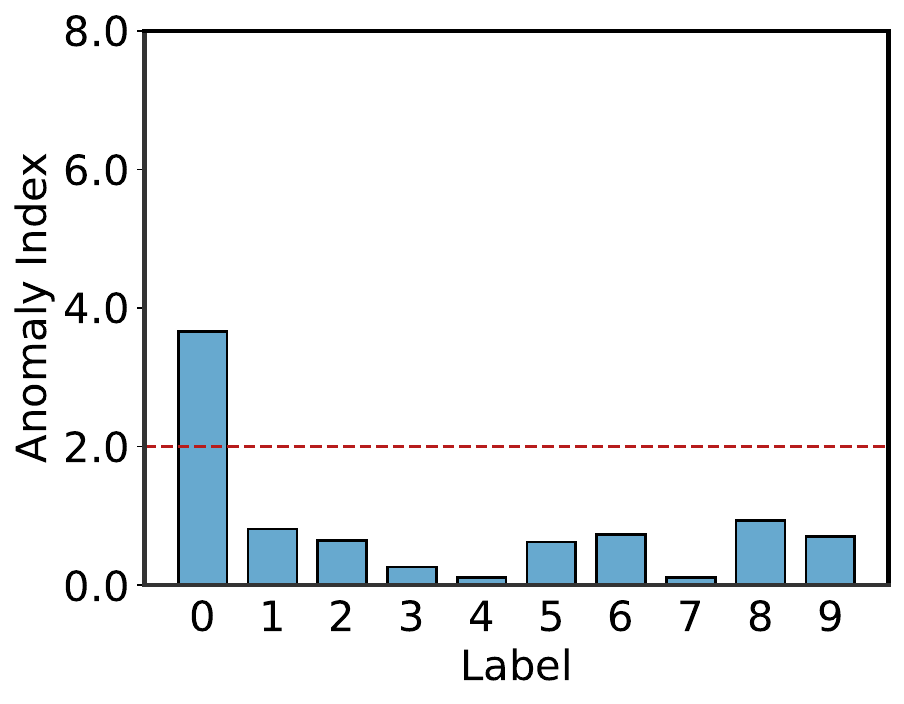}
        \caption{LF}
        \label{fig4-4}
    \end{subfigure}
  \hfill
  \begin{subfigure}[b]{0.24\linewidth}
        \centering
        \includegraphics[width=\columnwidth]{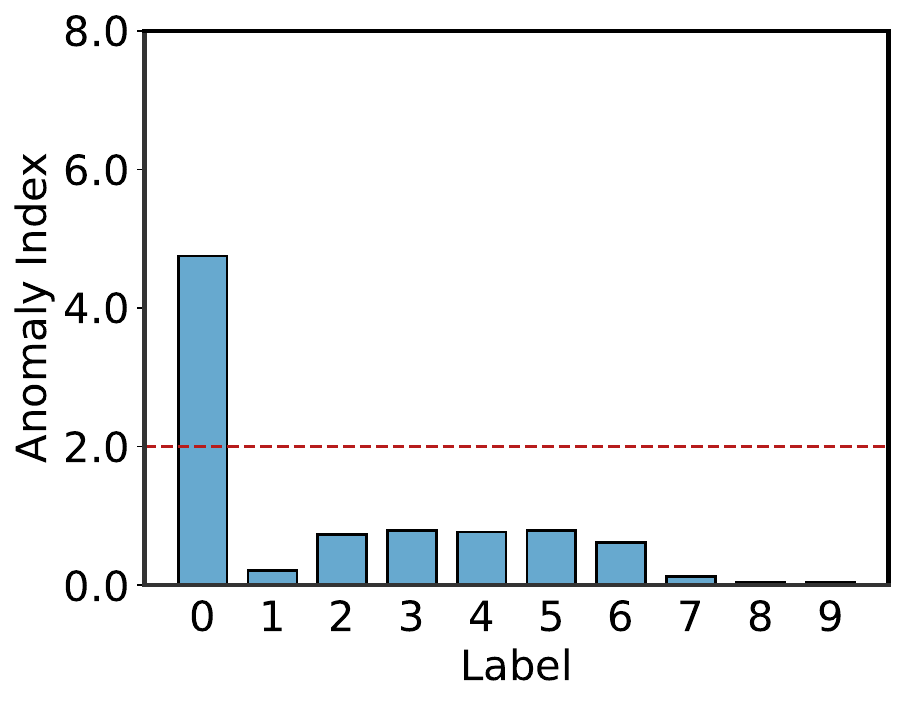}
        \caption{LC}
        \label{fig4-5}
    \end{subfigure}
  \hfill
  \begin{subfigure}[b]{0.24\linewidth}
        \centering
        \includegraphics[width=\columnwidth]{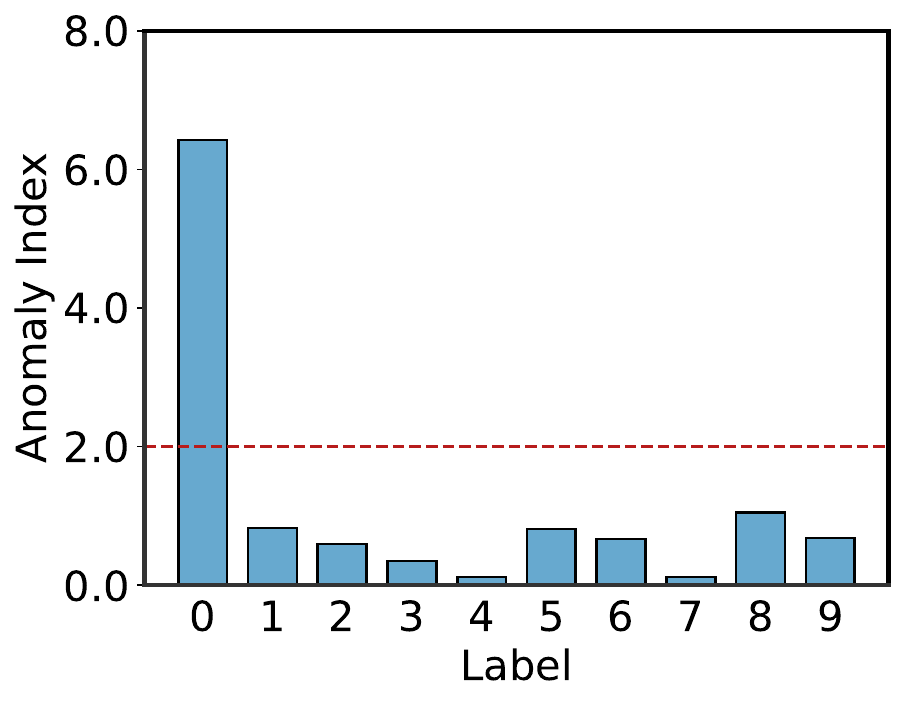}
        \caption{WaNet}
        \label{fig4-6}
    \end{subfigure}
  \hfill
  \begin{subfigure}[b]{0.24\linewidth}
        \centering
        \includegraphics[width=\columnwidth]{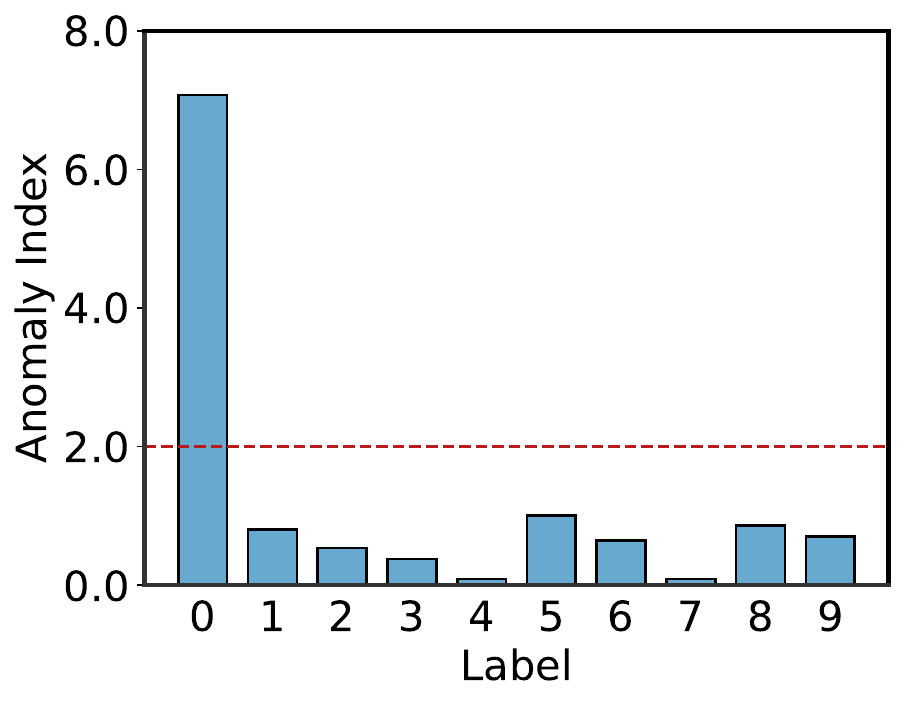}
        \caption{IA}
        \label{fig4-7}
    \end{subfigure}
  \hfill
  \begin{subfigure}[b]{0.24\linewidth}
        \centering
        \includegraphics[width=\columnwidth]{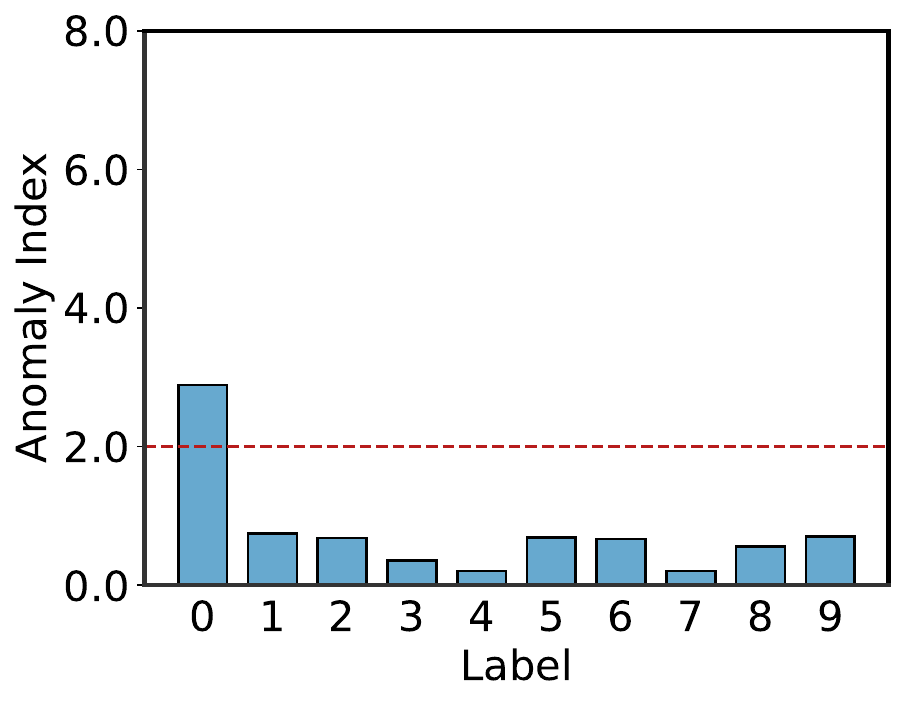}
        \caption{LIRA}
        \label{fig4-8}
    \end{subfigure}
    \caption{Target label detection on poisoned datasets.}
    \label{fig4}
\end{figure*}

\subsection{Comparison with Input Purification}
\label{subsec:comparison with input purification methods}

Furthermore, Table~\ref{tab7} shows the comparison results between \tool~and three input purification methods on CIFAR10. We utilized the official PyTorch implementation of Februus on Github\footnote{\url{https://github.com/AdelaideAuto-IDLab/Februus}}. While for ZIP and Sancdifi, we conducted experiments on \tool~according to the settings described in their respective papers, and then compared our results with the performance reported in their papers. Specifically, for ZIP, we conducted experiments against BadNets and Blended attacks using ResNet34, while for Sancdifi, we employed ResNet50 against BadNets. As Sancdifi only reports CAR (Clean Accuracy Reduction) in its paper, we present CAR instead of ACC in Table~\ref{tab7}. The left side showcases the performance of other defense methods, while the right side displays the defense results of \tool~under the same experimental settings. Notably, \tool~exhibits the best defense performance, achieving an average increase of 0.31\% in CAR and the lowest ASR of 1.27\%.

\begin{table}[ht]
\centering
\footnotesize
\tabcolsep=0.09cm
\renewcommand\arraystretch{1.5}
    \begin{tabular}{cccc}
    \hline
    \textbf{Defense} & \textbf{Attack} & \textbf{CAR} & \textbf{ASR}\\
    \hline
    \multirow{8}{*}{\shortstack{\textbf{Februus}\\\textbf{(PreAct-ResNet18)}}} & \textbf{BadNets}	& -22.06/\textbf{+0.85} & 99.23/\textbf{0.73} \\
     & \textbf{Blended}	& -20.77/\textbf{-0.07}	& 89.39/\textbf{1.11}\\
     & \textbf{SSBA}	& -21.31/\textbf{+0.20}	& 43.40/\textbf{1.31}\\
     & \textbf{LF}	& -21.06/\textbf{+0.71}	& 96.76/\textbf{4.61}\\
     & \textbf{LC}	& -20.22/\textbf{-0.48}	& 97.30/\textbf{1.91}\\	
     & \textbf{WaNet}	& -25.49/\textbf{-0.19}	& 26.84/\textbf{1.00}\\		
     & \textbf{IA} & -19.61/\textbf{+1.04} & 62.99/\textbf{0.58}\\	& \textbf{LIRA}	& -20.62/\textbf{+0.01}	& 98.74/\textbf{0.80}\\
     \hline
     \multirow{2}{*}{\shortstack{\textbf{ZIP}\\\textbf{(ResNet34)}}} & \textbf{BadNets}	& -8.00/\textbf{+0.86}	& 1.30/\textbf{0.56}\\
      & \textbf{Blended}	& -5.42/\textbf{+0.15}	& 3.29/\textbf{0.73}\\
    \hline
    \textbf{Sancdifi}\textbf{(ResNet50)} & \textbf{BadNets}	& -2.00/\textbf{+0.37}	& 12.00/\textbf{0.62}\\
    \hline
    \end{tabular}
\caption{Comparison between input purification methods (left) and \tool~(right).}
\label{tab7}
\end{table}

\section{Performance on Target Label Detection}
\label{sec:performance on target label detection}

\begin{figure}[htbp]
   \centering
    \includegraphics[width=0.6\columnwidth]{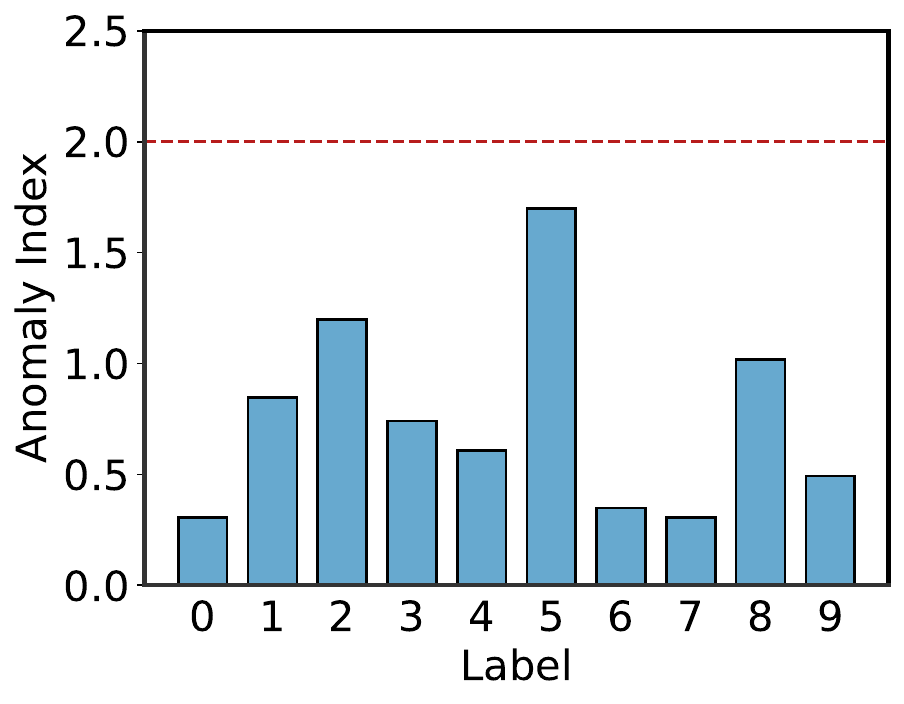}
    \caption{Target label detection on benign dataset.}
    \label{fig5}
\end{figure}

Figure~\ref{fig4} presents the results of target label detection on CIFAR10. Across the poisoned datasets created using eight attack methods, the average anomaly index for the target label is 5.4323, significantly surpassing the threshold value of 2. Moreover, \tool~produces no false positives for benign labels within these poisoned datasets, with an average maximum anomaly index of 1.0010 for benign labels. These results further highlight the effectiveness of \tool~in target label detection. Additionally, we evaluate the performance of \tool~on the benign dataset. As shown in Figure~\ref{fig5}, \tool~also exhibits no false positives for benign labels. The maximum anomaly index is 1.6992, further validating the reliability of \tool~in determining whether a dataset has been poisoned.

\begin{table*}[ht]
\centering
\footnotesize
\tabcolsep=0.084cm
\renewcommand\arraystretch{1.5}
\begin{tabular}{cccccccccccccccc}
    \hline
    \textbf{Dataset} & \multicolumn{8}{c}{\textbf{CIFAR10}} & \multicolumn{7}{c}{\textbf{Tiny ImageNet}}\\
    \cmidrule(lr){1-1} \cmidrule(lr){2-9}\cmidrule(lr){10-16}
    \textbf{Attack} & \textbf{BadNets} & \textbf{Blended} & \textbf{SSBA} & \textbf{LF} & \textbf{LC} & \textbf{WaNet} & \textbf{IA} & \textbf{LIRA} & \textbf{BadNets} & \textbf{Blended} & \textbf{SSBA} & \textbf{LF} & \textbf{Ftrojan} & \textbf{WaNet} & \textbf{IA}\\
    \hline
    \textbf{DBD} & 81.76 & 84.17 & 82.14 & 81.87 & \textbf{99.18} & 84.17 & 82.53 & 83.35 & 87.77 & 85.06 & 88.51 & 87.40 & 88.38 & 84.89 & 88.60\\
    \textbf{\tool} & \textbf{98.74} & \textbf{98.07} & \textbf{98.27} & \textbf{97.95} & 96.15 & \textbf{98.14} & \textbf{98.50} & \textbf{99.00} &  \textbf{96.07} & \textbf{95.59} & \textbf{89.90} & \textbf{95.42} & \textbf{96.21} & \textbf{96.06} & \textbf{95.78}\\
    \hline
\end{tabular}
\caption{Accuracy of labels in purified datasets compared to DBD.}
\label{tab8}
\end{table*}

\begin{table*}[ht]
\centering
\footnotesize
\tabcolsep=0.09cm
\renewcommand\arraystretch{1.5}
\begin{tabular}{cccccccccccccccc}
    \hline
    \textbf{Dataset} & \multicolumn{8}{c}{\textbf{CIFAR10}} & \multicolumn{7}{c}{\textbf{Tiny ImageNet}}\\
    \cmidrule(lr){1-1} \cmidrule(lr){2-9}\cmidrule(lr){10-16}
    \textbf{Attack} & \textbf{BadNets} & \textbf{Blended} & \textbf{SSBA} & \textbf{LF} & \textbf{LC} & \textbf{WaNet} & \textbf{IA} & \textbf{LIRA} & \textbf{BadNets} & \textbf{Blended} & \textbf{SSBA} & \textbf{LF} & \textbf{Ftrojan} & \textbf{WaNet} & \textbf{IA}\\
    \hline
    \textbf{FID} & 0.63 & 0.99 & 0.88 & 1.09 & 0.29 & 0.70 & 0.64 & 0.46 &  0.24 & 0.27 & 0.25 & 0.31 & 0.29 & 0.24 & 0.88 \\
    \textbf{IS} & 10.01 & 9.95 & 9.97 & 9.91 & 10.13 & 9.93 & 9.96 & 10.06 &  28.80 & 28.56 & 28.24 &  28.91 & 28.75 & 28.74 & 28.32 \\
    \hline
\end{tabular}
\caption{Quality of images in purified datasets.}
\label{tab9}
\end{table*} 

\section{Evaluation on Purified Dataset}
\label{sec:evaluation on purified dataset}

\noindent\textbf{Label Accuracy.} Table~\ref{tab8} shows the label accuracy of the purified dataset in contrast to the benign dataset. We compare the performance of \tool~with DBD, which guesses the ground-truth labels of suspicious samples using MixMatch during the semi-supervised fine-tuning process. Our label accuracy results outperform DBD, with an average of 98.10\% on CIFAR10 and 95.00\% on Tiny ImageNet, compared to the 84.90\% on CIFAR10 and 87.23\% on Tiny ImageNet of DBD. These results highlight the efficacy of \tool~in accurately determining ground-truth labels. 

\noindent\textbf{Image Quality.} We further evaluate the quality of images in the purified datasets using \textit{Fréchet Inception Distance} (FID)~\cite{heusel2017gans} and \textit{Inception Score} (IS)~\cite{salimans2016improved} metrics. FID measures the distribution similarity between the images in the purified dataset and those in the benign dataset, while IS assesses the quality and diversity of images in the purified dataset. A lower FID indicates a closer alignment in distribution, while a higher IS indicates superior quality and diversity of images. As shown in Table~\ref{tab9}, the purified dataset exhibits a high similarity in distribution with the benign dataset, with an average FID of 0.71 on CIFAR10 and 0.35 on Tiny ImageNet. Moreover, the images in the purified dataset demonstrate comparable quality and diversity to the original dataset, with an average IS of 9.99 on CIFAR10 and 28.62 on Tiny ImageNet, while the IS values for the original CIFAR10 and Tiny ImageNet are 10.24 and 29.77, respectively. These results further validate the effectiveness of \tool~in constructing a high-quality purified dataset. 

\begin{figure*}[!t]
   \centering
    \includegraphics[width=2\columnwidth]{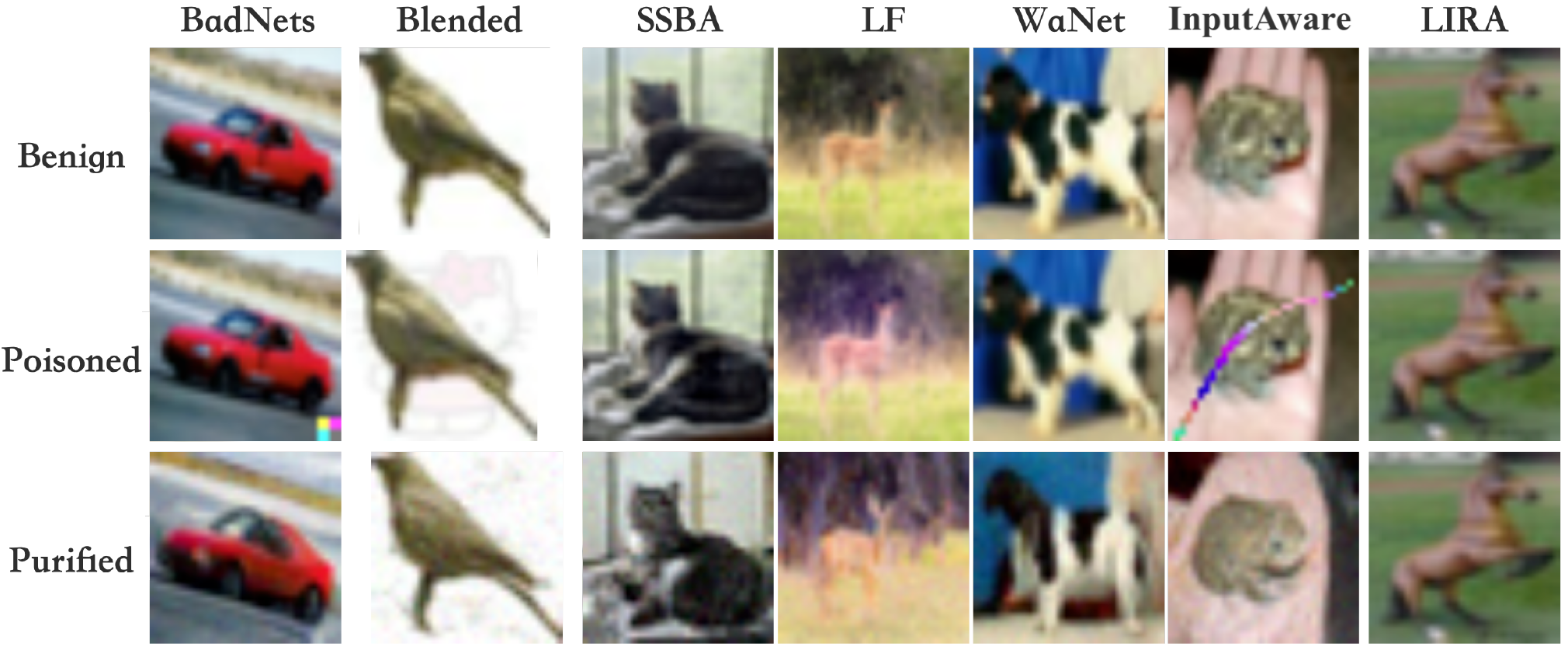}
    \caption{Benign images (first row), poisoned images (second row) and purified images (third row) from CIFAR10.}
    \label{fig6}
\end{figure*}

\begin{figure*}[!t]
   \centering
    \includegraphics[width=2\columnwidth]{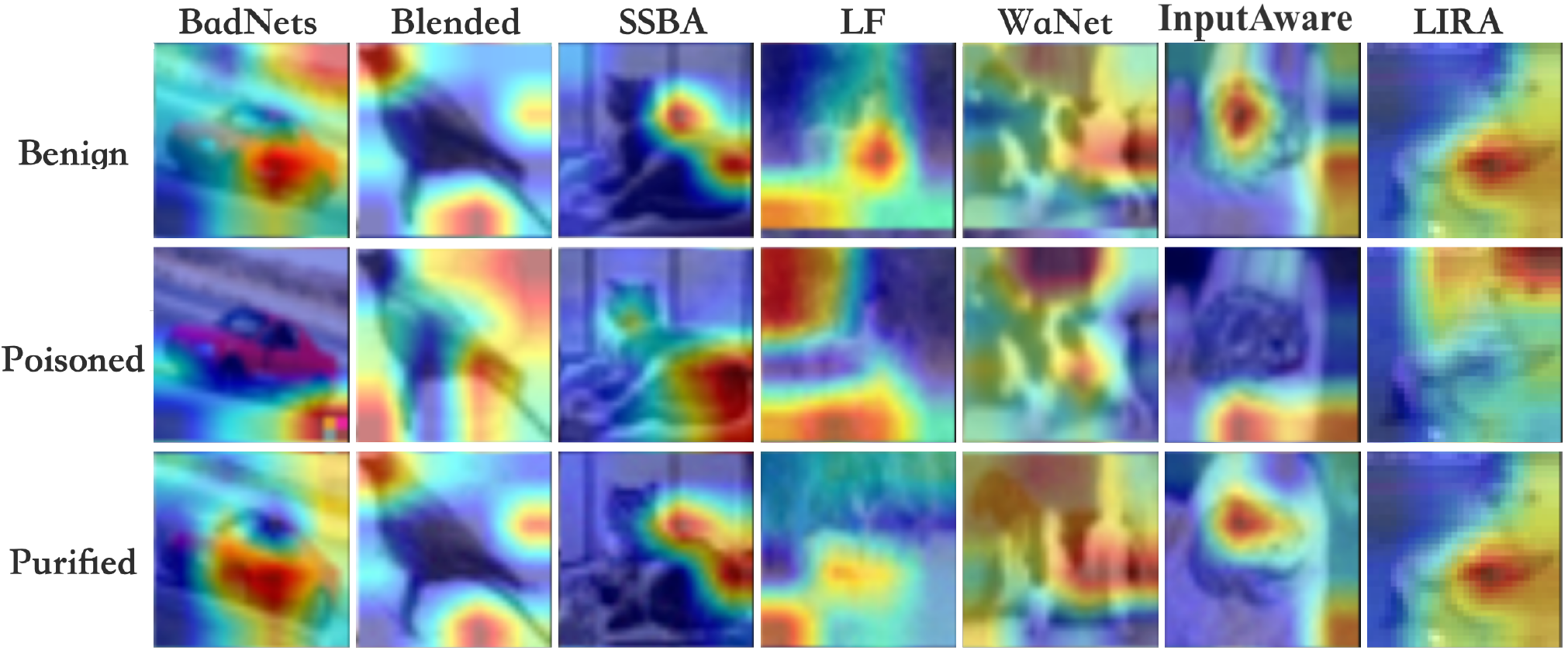}
    \caption{GradCAM visualization for benign images (first row), poisoned images (second row) and purified images (third row).}
    \label{fig7}
\end{figure*}

Moreover, Figure~\ref{fig6} presents the benign images (first row), poisoned images (second row) and purified images (third row) from CIFAR10. Notably, \tool~demonstrates its capability to effectively eliminate the trigger features from the poisoned images while preserving the benign features intact. This observation is further supported by the GradCAM Visualization of these images as shown in Figure~\ref{fig7}. These results provide strong evidence of the effectiveness of \tool~in the purification of poisoned images. Figure~\ref{fig6} and Figure~\ref{fig7}
further elucidate our motivation for using diffusion models as well as the effectiveness of the forward and reverse process which also shown in Figure~\ref{fig1}. While training on the poisoned dataset, the backdoor model is overfitting to the trigger features, which gain more attention than benign features, thus resulting in misclassification of poisoned images into the target label. The forward process eliminates trigger features through the introduction of noise, while the reverse process restores the degraded benign features. As a result, in purified images, the effective elimination of trigger features enables the benign features to regain the model’s focus, allowing the model to correctly identify purified images as their ground-truth labels.

\noindent\textbf{Performance of Newly Trained Model.} Furthermore, Table~\ref{tab10} shows the performance of a newly trained model on the purified dataset (different from the benign model obtained via \tool), with an average ACC and ASR of 93.05\% and 1.13\% on CIFAR10 and 65.75\% and 0.41\% on Tiny ImageNet. These results highlight the reusability of the purified datasets, which can benefit an unlimited number of model training tasks without any concern for reduced benign accuracy or the potential injection of backdoors.

\begin{table*}[ht]
\centering
\footnotesize
\tabcolsep=0.09cm
\renewcommand\arraystretch{1.5}
\begin{tabular}{cccccccccccccccc}
    \hline
    \textbf{Dataset} & \multicolumn{8}{c}{\textbf{CIFAR10}} & \multicolumn{7}{c}{\textbf{Tiny ImageNet}}\\
    \cmidrule(lr){1-1} \cmidrule(lr){2-9}\cmidrule(lr){10-16}
    \textbf{Attack} & \textbf{BadNets} & \textbf{Blended} & \textbf{SSBA} & \textbf{LF} & \textbf{LC} & \textbf{WaNet} & \textbf{IA} & \textbf{LIRA} & \textbf{BadNets} & \textbf{Blended} & \textbf{SSBA} & \textbf{LF} & \textbf{Ftrojan} & \textbf{WaNet} & \textbf{IA}\\
    \hline
    \textbf{ACC} & 93.47 & 92.83 & 93.16 & 92.84 & 93.37 & 92.40 & 92.87 & 93.49 & 66.34 & 66.08 & 66.34 & 65.83 & 64.43 & 65.44 & 65.81\\ 
    \textbf{ASR} & 0.81 & 0.76 & 1.03 & 3.44 & 1.91 & 0.67 & 0.39 & 0.77  & 0.12 & 0.25 & 0.34 & 0.44 & 1.34 & 0.19 & 0.20 \\
    \hline
\end{tabular}
\caption{Performance of models newly trained on purified datasets.}
\label{tab10}
\end{table*}

\section{Defense against Various Attack Scenarios}
\label{sec:defense against various attack scenarios}

We assess \tool~against various attack scenarios using datasets poisoned with different poison rates, target label numbers and target labels. Our evaluations are conducted on CIFAR10, using BadNets as a case study. 

\begin{figure}[htbp]
  \centering
  \begin{subfigure}[b]{0.32\linewidth}
        \centering
        \includegraphics[width=\columnwidth]{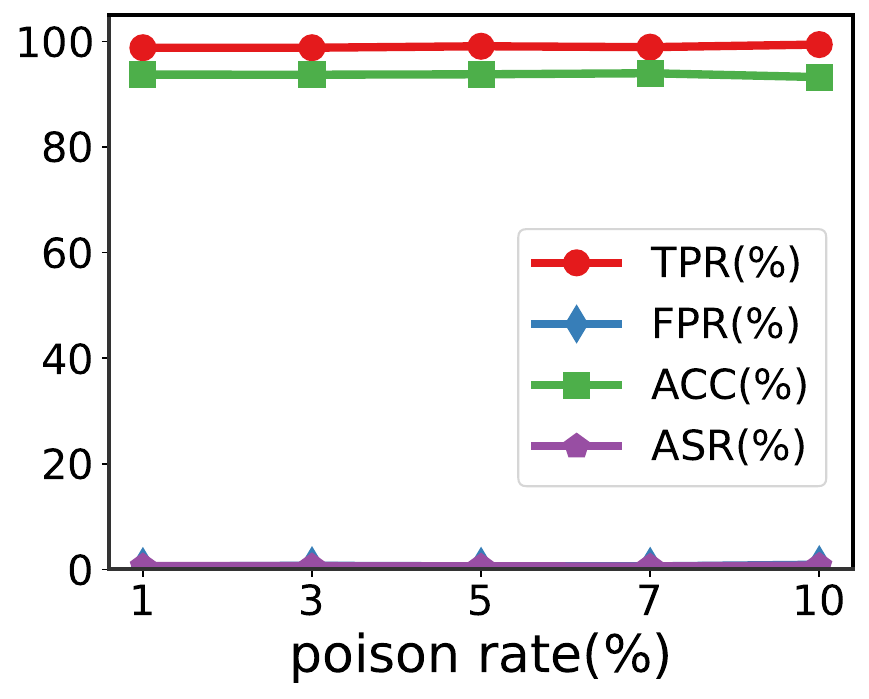}
        \caption{poison rates}
        \label{fig8-1}
  \end{subfigure}
  \hfill
  \begin{subfigure}[b]{0.32\linewidth}
    \centering
    \includegraphics[width=\columnwidth]{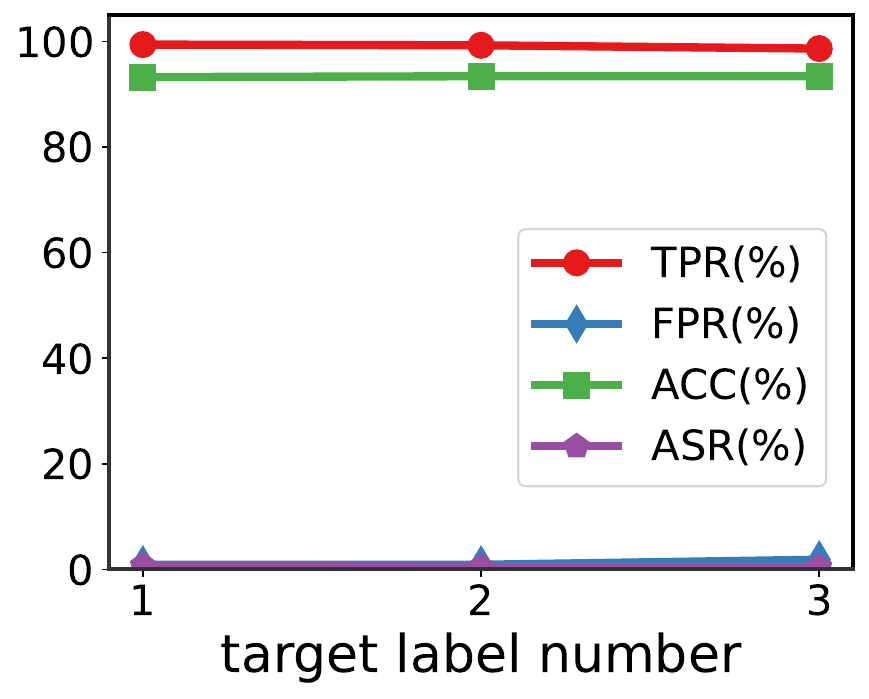}
    \caption{numbers}
    \label{fig8-2}
  \end{subfigure}
  \hfill
  \begin{subfigure}[b]{0.32\linewidth}
    \centering
    \includegraphics[width=\columnwidth]{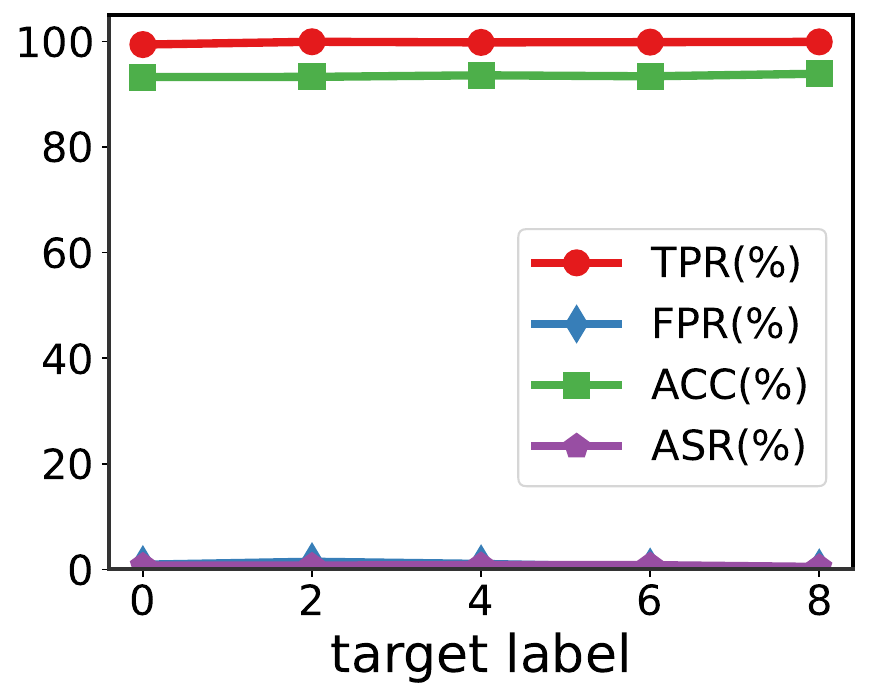}
    \caption{target labels}
    \label{fig8-3}
  \end{subfigure}
  \caption{Defense performance in various attack scenarios.}
  \label{fig8}
\end{figure}

\noindent\textbf{Different Poison Rates.} We consider poison rates varying in $\{1\%, 3\%, 5\%, 7\%, 10\%\}$. It is worth noting that accurately detecting the target label in datasets with low poison rates (\eg, 1\%) is challenging, whereas effectively purifying all poisoned samples in datasets with high poison rates (\eg, 10\%) is difficult. As shown in Figure~\ref{fig8-1}, \tool~demonstrates remarkable performance across varying poison rates, achieving the ASR of no more than 0.73\%, and ACC higher than 93.24\%. Additionally, the FPR remains below 0.89\%, with TPR surpassing 98.80\%. These results highlight the robustness of \tool~against various numbers of poisoned samples in datasets.

\begin{figure}[htbp]
  \centering
  \begin{subfigure}[b]{0.18\linewidth}
        \centering
        \includegraphics[width=\columnwidth]{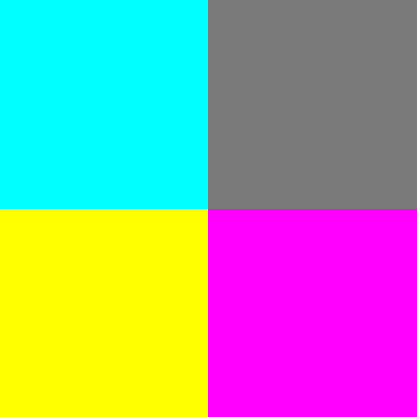}
        \caption{Label 0}
        \label{fig9-1}
  \end{subfigure}
  \hspace{0.75cm}
  \begin{subfigure}[b]{0.18\linewidth}
    \centering
    \includegraphics[width=\columnwidth]{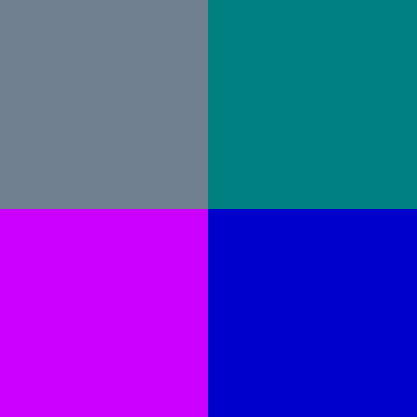}
    \caption{Label 1}
    \label{fig9-2}
  \end{subfigure}
  \hspace{0.75cm}
  \begin{subfigure}[b]{0.18\linewidth}
        \centering
        \includegraphics[width=\columnwidth]{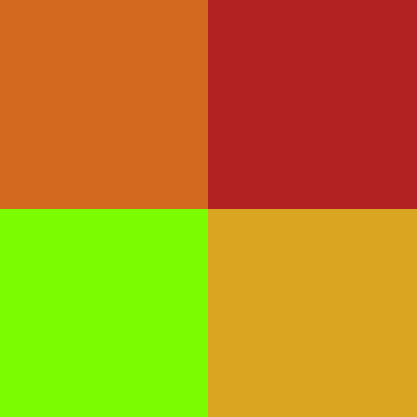}
        \caption{Label 2}
        \label{fig9-3}
  \end{subfigure}
  \caption{Triggers for different target labels. Each trigger consists of $4 \times 4$ pixels and is positioned in the bottom right corner of the image.}
  \label{fig9}
\end{figure}

\noindent\textbf{Different Target Label Numbers.} We conduct experiments on datasets with target label numbers varying in $\{1, 2, 3\}$. The specific triggers for different target labels are shown in Figure~\ref{fig9}. The overall poison rate is fixed at 10\%, which means that if there are two target labels, the poison rate for each target label accounts for 5\%. As shown in Figure~\ref{fig8-2}, \tool~demonstrates robust performance across varying target label numbers, achieving TPR and ACC consistently above 98.64\% and 93.24\%, respectively, while maintaining FPR and ASR below 1.83\% and 0.73\%. Moreover, our target label detection method proves to be highly effective, detecting all target labels in all poisoned datasets while avoiding any false positives for benign labels.

\noindent\textbf{Different Target Labels.} We conduct experiments on poisoned datasets with the target label being $\{0, 2, 4, 6, 8\}$ to assess the impact of target label selection on the performance of \tool. As shown in Figure~\ref{fig8-3}, \tool~exhibits robust performance against different target labels, achieving TPR and ACC consistently above 98.58\% and 93.24\%, respectively, while maintaining FPR and ASR below 1.45\% and 0.82\%.

\section{More Ablation Studies}
\label{sec:more ablation studies}

We evaluate the impacts of our proposed purified image selection strategy and ground-truth label determination strategy on the performance of \tool. All experiments are conducted on CIFAR10 against BadNets.

\begin{figure}[htbp]
  \centering
  \begin{subfigure}[b]{0.48\linewidth}
        \centering
        \includegraphics[width=\columnwidth]{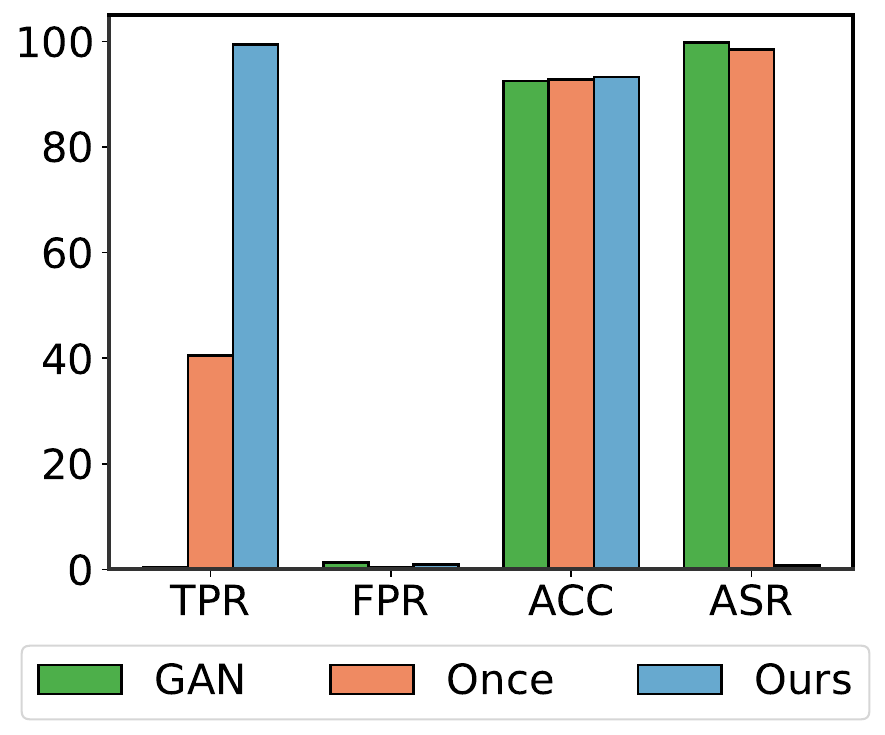}
        \caption{Purified image}
        \label{fig10-1}
  \end{subfigure}
  \hfill
  \begin{subfigure}[b]{0.48\linewidth}
        \centering
        \includegraphics[width=\columnwidth]{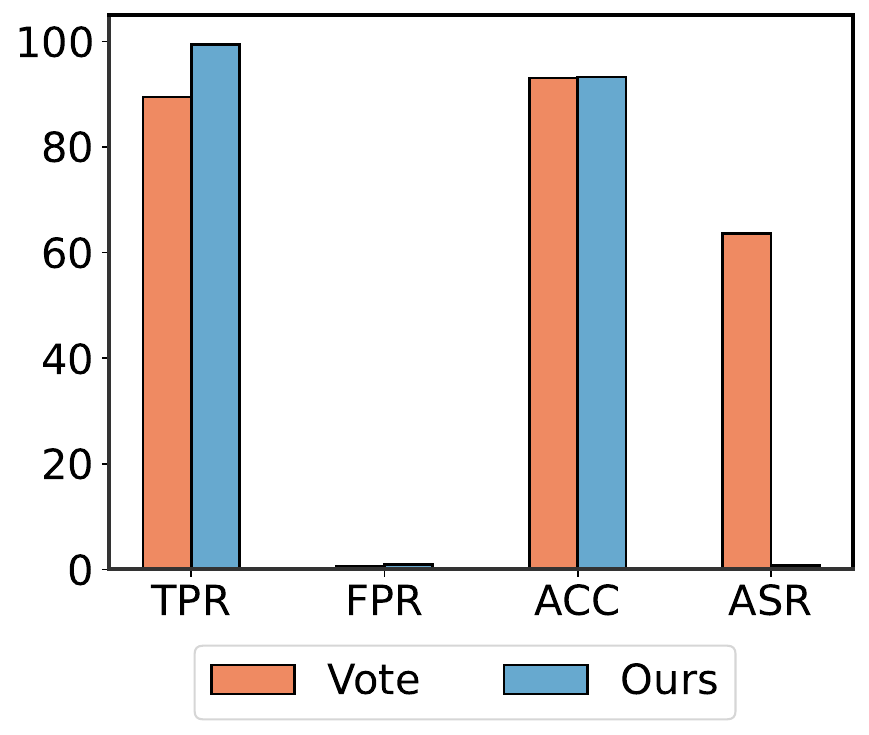}
        \caption{Ground-truth label}
        \label{fig10-2}
  \end{subfigure}
  \caption{Impact of our proposed purified image selection and ground-truth label determination strategy.}
  \label{fig10}
\end{figure}

\noindent\textbf{Purified Image Selection.} Our purified image selection strategy involves using the original image as the purified image for samples in the Benign Set. While for samples in the Poisoned Set and Suspicious Set, we select their purified images from their respective candidate sets by considering both pixel and feature distance using Equation~\ref{eq5}. We compare our strategy with the baseline approach commonly used in input purification~\cite{shi2023blackbox,may2023salient} and adversarial purification methods during the inference stage~\cite{yoon2021adversarial,nie2022diffusion,xiao2023densepure,carlini2023certified}, which performs the forward and reverse process of diffusion models once and uses the resulting image as the purified image. In addition, we also utilize the GAN employed in Februus~\cite{doan2020februus} to generate purified images for comparison. As shown in Figure~\ref{fig10-1}, the once-based strategy only detects 40.52\% of the poisoned samples, leading to a high ASR of 98.44\%. This result indicates that the technique for using diffusion models in input purification and adversarial purification during the inference stage is insufficient to purify the poisoned training dataset. While the TPR of the GAN-based strategy is only 0.40\%, which means a considerable number of poisoned samples remain in the purified dataset, resulting in a high ASR of 99.77\%. In contrast, \tool~achieves a high TPR of 99.40\% and a remarkably low ASR of only 0.73\%, thereby validating the effectiveness of our strategy. 

\noindent\textbf{Ground-Truth Label Determination.} Our ground-truth label determination strategy involves using the consensus label in the candidate set as the ground-truth label for the samples in the Benign Set and Poisoned Set ($y_i$ for $B$ and $y_g$ for $P$ in Equation~\ref{eq3}). While for samples in the Suspicious Set, a benign model is trained on the available purified dataset to determine their ground-truth labels. Alternatively, it is straightforward for the defender to use a voting-based strategy that does not distinguish the samples into the above three sets, but directly votes on the labels in their respective candidate sets and uses the most occurrence as their ground-truth labels, and we consider such voting-based strategy as our baseline for comparison. The defense performance of these two strategies is shown in Figure~\ref{fig10-2}. Note that while the voting-based strategy can identify 89.48\% of the poisoned samples, the remaining 10.52\% samples can still successfully inject a backdoor into the victim model with 63.63\% ASR, indicating the challenge of effectively purifying the poisoned training dataset. In contrast, the ASR of the model trained on the dataset purified by \tool~is only 0.73\%, demonstrating the effectiveness of our strategy.

\end{document}